\documentclass[achemso,aip,jcp,preprint,superscriptaddress,floatfix]{revtex4-1}
\usepackage{rotating}
\usepackage{graphicx}
\usepackage{xcolor}
\usepackage{mathtools}
\usepackage{multirow}
\usepackage{amsmath}
\usepackage{amssymb}
\usepackage{tikz}
\usepackage{threeparttable}

\usepackage{color}
\usepackage{subfigure}
\usepackage{palatino}
\usepackage{mathpazo}
\usepackage{makecell}
\usepackage{overpic}
\usepackage{verbatim}
\usepackage{listings}
\usepackage{tcolorbox}
\usepackage{framed}
\usepackage{float}

\usepackage{url}

\usetikzlibrary{shapes.geometric, arrows}
\usetikzlibrary{calc}
\usetikzlibrary{decorations.pathmorphing}  
\usetikzlibrary{fit}					   
\usetikzlibrary{backgrounds}

\usepackage{xargs} 
\usepackage{lipsum}  
\usepackage[colorinlistoftodos,prependcaption,textsize=tiny]{todonotes}

\newcommandx{\addcite}[2][1=]{\todo[linecolor=red,backgroundcolor=red!25,bordercolor=red,#1]{#2}}
\newcommandx{\checkthis}[2][1=]{\todo[linecolor=blue,backgroundcolor=blue!25,bordercolor=blue,#1]{#2}}
\newcommandx{\addsome}[2][1=]{\todo[linecolor=green,backgroundcolor=green!25,bordercolor=green,#1]{#2}}
\newcommandx{\modifythis}[2][1=]{\todo[linecolor=cyan,backgroundcolor=cyan!25,bordercolor=cyan,#1]{#2}}

\begin{document}
\title[]{A Forecasting System of Computational Time of DFT/TDDFT Calculations under the Multiverse ansatz via Machine Learning and Cheminformatics}
\author{Shuo Ma}
\affiliation{Computer Network Information Center, Chinese Academy of Sciences, Beijing 100190, China}
\affiliation{School of Computer Science and Technology, University of Chinese Academy of Sciences, Beijing 101408, China}
\author{Yingjin Ma}
\email{yingjin.ma@sccas.cn}
\affiliation{Computer Network Information Center, Chinese Academy of Sciences, Beijing 100190, China}
\affiliation{Center of Scientific Computing Applications \& Research, Chinese Academy of Sciences, Beijing 100190, China}
\author{Baohua Zhang}
\affiliation{Computer Network Information Center, Chinese Academy of Sciences, Beijing 100190, China}
\affiliation{Center of Scientific Computing Applications \& Research, Chinese Academy of Sciences, Beijing 100190, China}
\author{Yingqi Tian}
\affiliation{Computer Network Information Center, Chinese Academy of Sciences, Beijing 100190, China}
\affiliation{School of Computer Science and Technology, University of Chinese Academy of Sciences, Beijing 101408, China}
\author{Zhong Jin}
\email{zjin@sccas.cn}
\affiliation{Computer Network Information Center, Chinese Academy of Sciences, Beijing 100190, China}
\affiliation{Center of Scientific Computing Applications \& Research, Chinese Academy of Sciences, Beijing 100190, China}

\begin{abstract}
With the view to a better performance in task assignment and load-balancing,
a top-level designed forecasting system for predicting computational times of density-functional theory (DFT)/time-dependent density-functional theory (TDDFT) calculations is presented. {The computational time is assumed as the intrinsic property for the molecule. 
Basing on this assumption, the forecasting system is established using the "reinforced concrete", which combines the cheminformatics, several machine-learning (ML) models, and the framework of many-world interpretation (MWI) in multiverse ansatz. 
Herein, the cheminformatics is used to recognize the topological structure of molecules, the ML models are used to build the relationships between topology and computational cost, and the MWI framework is used to hold various combinations of DFT functionals and basis sets in DFT/TDDFT calculations. 
Calculated results of molecules from DrugBank dataset show that 1) it can give quantitative predictions of computational costs, typical mean relative errors can be less than 0.2 for DFT/TDDFT calculations with derivations of $\pm25\%$ using the exactly pre-trained ML models, 2) it can also be employed to various combinations of DFT functional and basis set cases without exactly pre-trained ML models, while only slightly enlarge predicting errors.
}

\end{abstract}

\maketitle

\section{Introduction}

\textit{Ab initio} electronic structure methods {are becoming} more and more popular in the chemistry community, {as it has been reported} that the \textit{ab initio} methods illustrate the chemical mechanism in its original view, {i.e., at the electron} level. \cite{Friesner6648, HERGERT2016165, ALDOURI201533, scerri1994has, bash1996progress, zhao2019chemical}
Normally, one needs to consider the computational costs of \textit{ab initio} (i.e. first principle) methods when determining whether they are appropriate for the problem at hand {or not. As shown in Fig.\ref{scales},} when compared to much less accurate approaches, such as molecular mechanics, \textit{ab initio} methods often take larger amounts of computer time, memory, and disk space due to their scales. {For example, the Hartree-Fock self-consistent field (HFSCF) and density functional theory (DFT) already shows scales in the range from N$^{2.x}$ to N$^{4.x}$ with N is the system magnitude parameters, not specificly the number of basis functions. \cite{jensen2017introduction, helgaker2014molecular} The quantitative solutions (i.e. electronic correlation approaches) like M\o ller-Plesset perturbation theory up to 2nd order (MP2)\cite{PhysRev.46.618}, coupled cluster approaches with single and double excitations (CCSD)\cite{Purvis1982,cullen1982} and iterative perturbed treatment (CCSD(T)) increase the scales by two or more orders, respectively. 
In general, the dynamic correlations can be well accounted by the aboved-mentioned DFT, MP2, and CCSD/CCSD(T) approaches. However, the static correlations, i.e. quasi-degeneracy, are normally accounted by the multi-configurational (MC) and multi-reference (MR) approaches.\cite{shepard2012} The scales of MC/MR approaches can be in a large range due to the so-called active space, in which the super-position of all possible configurations can be took into account for describing the quasi-degeneracy. \cite{helgaker2014molecular, shepard2012}

\begin{figure}[hbp] 
	\centering 
	\includegraphics[scale=1.60,bb=25 0 375 250]{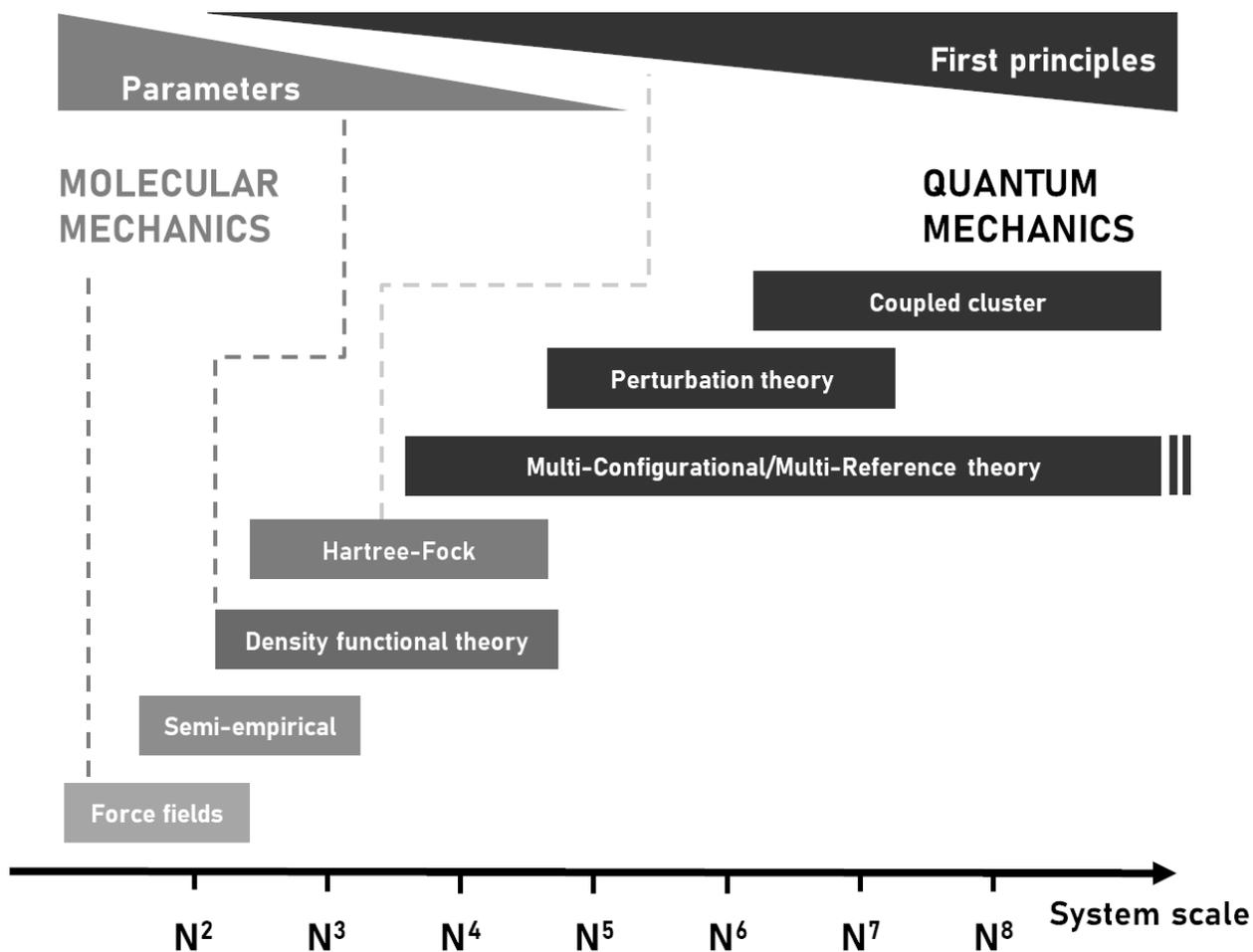}
	\caption{The system scales under different simulating approaches.}
	\label{scales}
\end{figure} 

{{Normally, one needs to choose the simulating approaches carefully by considering the studied systems as well as the desired accuracy. 
Apparently, the choice of simulating approaches, or even the choice of parameters in a given approach is also dominated by the computational resources. \cite{helgaker2014molecular}
However, the predictable pace of Moore's law \cite{brock2006understanding} cannot easily compensate for the difficulties caused by the conflict between computational resources and computational scales. 
Thus, there are actual demands for predicting the possible computational costs (e.g. time, memory, or disk space) before performing the simulations. 
It can benefit both researchers and computer centers in terms of computational resource utilization and computational resource scheduling, respectively.\cite{Heinen2019MachineLT}}} 
Additionally, a potential and specific usage can be expected in the high-throughput calculations using fragmentation approaches,\cite{gordon2012fragmentation, 2019Improved, wang2020combining, fedorov2009fragment, babu2003ab} in which the task assignment and load-balancing are highly demanded. Because the different fragments correspond to different computational costs that can not be well estimated before the practical calculations.  

{Several related works have been carried out in the area of computational cost modelling and job scheduling in the past decades. 
Most of the related works chose to make use of historical data and adopt methods of statistical analysis and machine learning (ML), with which the relationship between job features and the computational cost can be established. \cite{downey_predicting_1997, goos_using_1999, tsafrir_backfilling_2007, gaussier_improving_2015, sonmez_trace-based_2009, li_predicting_2016, negi_applying_2005, helmy_machine_2015, shulga_scheduling_2016, nadeem_optimizing_2013, singh_predicting_2007, matsunaga_use_2010}
One of the research prospects was characterized by laying emphasis on the meta information of jobs, including user types, number of processors applied for, execution time limits and so on. These works took the assumption that different jobs with similar meta information come with similar costs, which was validated by Downey\cite{downey_predicting_1997} and Smith's\cite{goos_using_1999} works. Time series approaches have been frequently used in this kind of works. 
For instance, 
Gaussier \textit{et al.}\cite{gaussier_improving_2015} designed a feature set including varieties of job history data as the input of a linear regression model. 
Moreover, many works focused on the the architectures of the targeted systems and use ML to make predictions. 
The training data usually come from hardware performance counters so that the states of computing components during executions can be considered. 
Li \textit{et al.}\cite{li_predicting_2016} used a support vector machine (SVM) model to predict the instruction number in each cycle and obtain the optimal thread mapping scheme. 
Helmy \textit{et al.}\cite{helmy_machine_2015} employed SVM, artificial neural network, and decision tree models to predict CPU burst time, and it can be further extended to a heterogeneous computing system as shown by Shulga and co-workers. \cite{shulga_scheduling_2016}
Apart from the above solutions targeting common programs, there are also some predictive methods specified for a certain type of programs. \cite{nadeem_optimizing_2013, singh_predicting_2007} 
It is worthy of note that Matsunaga \textit{et al.}\cite{matsunaga_use_2010} applied the PQR2 model to predict the execution time, memory, and disk occupations of two bioinformatics applications (BLAST \cite{altschul1990basic} and RAxML \cite{stamatakis2014raxml}). 
It has been reported that taking features that were highly correlated with application types, such as protein sequence length, yields much better results.} }

{{Regarding to} the field of computational chemistry, Papay \textit{et al.} {developed} a least square fitting method for graph-based component-wise runtime estimates in parallel SCF atomic computations in 1996. \cite{Papay1996PERFORMANCE} Antony \textit{et al.} used a linear model to simulate the runtime of SCF {algorithms} in {Gaussian applications} and to {estimate }the impacts of architectures in terms of the count of retired instructions and cache misses.\cite{Antony2011Modelling} 
Additionally, it is {noteworthy} that Mniszewski \textit{et al.} designed a class of tools for {predicting} the runtime of a molecular dynamics code,\cite{Mniszewski2015TADSim} allowing users to find the optimal combination of algorithmic methods and hardware options. 
However, as far as we know, there is nearly no related work concerning to the prediction of computation cost in the field of quantum chemistry, except {in the area of} quantum machine learning (QML) models that very recently introduced by Heinen and coworkers. \cite{Heinen2019MachineLT} 
They {demonstrated} that QML-based wall time predictions significantly improve job scheduling efficiency by {reducing} CPU time overhead ranging from 10\% to 90\%. 
{Until now, there is no universal solution for predicting the computational cost. 
The current QML solution is restricted to a specified computational approach and specified parameters, and training of a corresponding ML model is essential before practical predictions.\cite{Heinen2019MachineLT} 
However, it may not be convenient for training the specific model each time before the practical calculations. 
Thus, generalization ability should be one of the essential elements for an universal solution when predicting the computational cost. Additionally, the traditional ML models including the ensemble learning \cite{schuld2015}, recurrent \cite{biamonte2017} or graph-based neural network \cite{ciliberto2018}, may also be the good alternatives to the reported QML solutions.\cite{Heinen2019MachineLT}}

Herein, a top-level {design-based} forecasting framework is developed, {which aims} to {yield} reliable prediction {of} computational cost (mainly the computational time) with  a {high }degree of generalization ability.
{At this stage, we focus on its design and its confirmatory usage via the prediction of computational times of DFT/TDDFT single-point calculation.} 
In our design, cheminformatics is used to recognize the topological structure of molecules, and the ML models are used to establish the relation between topology and computational time. Additionally, the idea from many-world interpretation (MWI) in multiverse ansatz is used to gain generalization ability when treating with various combinations of DFT functional and basis sets, which are critical for practical calculations. \cite{baerends1997quantum, perdew2005prescription, cohen2012challenges}}

For the DFT calculations, the computational scaling is range from N$^{2.x}$ to N$^{4.x}$, as shown in Fig.\ref{scales} and detailed explained in appendix.
It {stems from the} evaluation of four-center two-electron repulsion integrals,
i.e.,
\begin{equation}\label{eq_2e}
    (\mu \nu | \lambda \sigma) = \int \int \phi_{\mu}(1) \phi_{\nu}(1) \frac{1}{r_{12}} \phi_{\lambda}(2) \phi_{\sigma}(2) d\tau_1 d\tau_2 , 
\end{equation}
where $\mu$, $\nu$, $\lambda$, and $\sigma$ {denote} indices of atomic orbitals. 
This scaling is the upper boundary for the {HF} or DFT calculations. \cite{strout1995quantitative, perez1997fast} 
However, many {two-electron} integrals are of negligible magnitude for large molecules, {and} some rigorous upper boundary conditions can be {applied} to the integrals. For {instance}, the Schwarz inequality \cite{steele2004cauchy}
\begin{equation}\label{eq_Schwarz}
    |(\mu \nu | \lambda \sigma)| \leqslant \sqrt{(\mu \nu | \mu \nu)(\lambda \sigma | \lambda \sigma)}
\end{equation}
can reduce the mathematical upper bounds {of all two-electron} integrals to be computed in an $N^2$ $logN$ process 
{by safely ignoring the predetermined negligible integrals. Additionally, larger molecular systems have a higher fraction of atomic orbitals sufficiently distant from each other to be considered non-interacting, thereby yielding negligible two-electron integrals and further decline the exponent for HF or DFT calculations. \cite{strout1995quantitative}} 

{
It was demonstrated by Strout and Scuseria that the number of basis functions $n$ can connect with the integrals via the scaling exponent $\alpha$
for the same molecular series,\cite{strout1995quantitative} e.g.
\begin{equation}\label{eq_alpha}
 (\frac{n_2}{n_1})^\alpha=(\frac{I_2}{I_1}),
\end{equation}
where \textit{n} denotes the number of basis functions and \textit{I} denotes the number of integrals.}
Because the computational cost of integrals are the upper boundary for the HF or DFT calculations, {such that a $t = a n^2 + b n +c $ type formula (exponent 2 is roughly approximated from $\alpha$)} can be expected as the working equation for rough prediction the time.
However, {the }simple polynomial equation or the exponential expression (Eq.\ref{eq_alpha}) is only suitable for the molecules in the same series and better without significant scale difference. When molecules {have} different spatial structures, the predicted results are normally too poor to be referred when using this type of regression {equation}. \cite{strout1995quantitative}
Additionally, it is not convenient for {this} regression analysis to consider multiple factors (e.g., basis number together with electrons, bond type, etc.), which should be considered when better predictions are needed. {Thus, the advantages of the proposed top-level design-based forecasting framework can be demonstrated in practical predictions.}

{The paper is organized as follows: in Sec. \uppercase\expandafter{\romannumeral2}, we present the design of the proposed framework, implementation details, and the workflow of the forecasting system; computational details presented in Sec. \uppercase\expandafter{\romannumeral3} while benchmark examples are presented in Sec. \uppercase\expandafter{\romannumeral4}; and finally, we draw the conclusions of the work in Sec. \uppercase\expandafter{\romannumeral5}.}
  

\section{Implementations}

\subsection{Chemical spaces containing the computational times} 

\begin{figure}[htbp]
	\includegraphics[scale=1.00, bb=50  0 450 220]{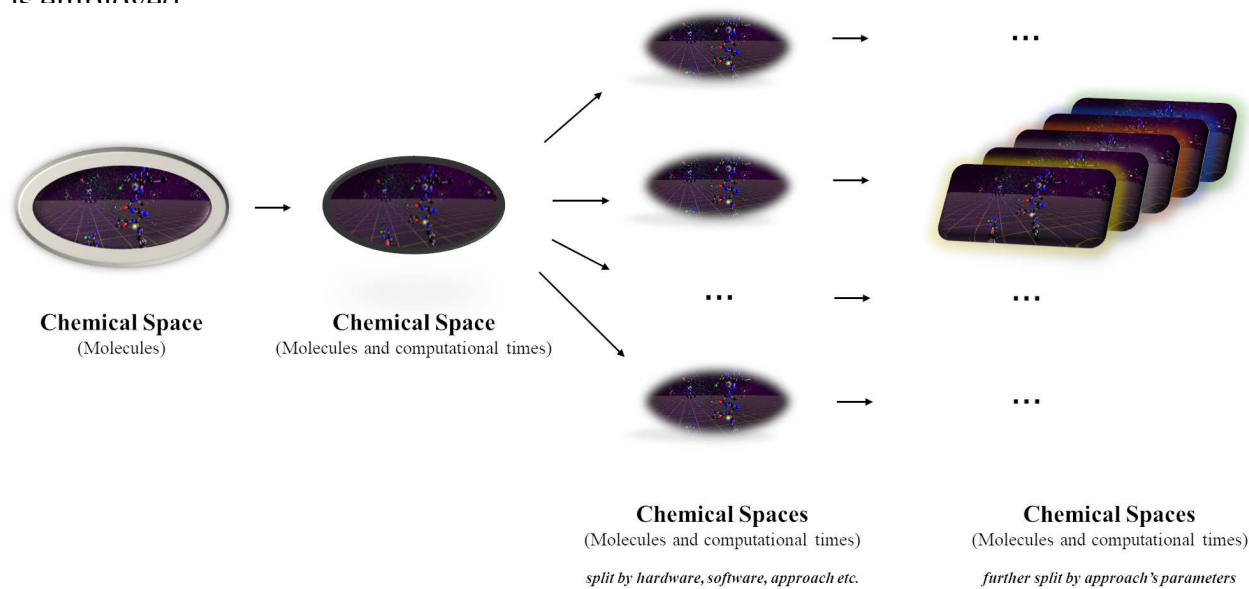}
	
	\caption{{The illustration of chemical spaces that contain molecular suits and computational times. The chemical spaces can be split by different computational schemes (e.g. hardware, software, and approaches, etc), and can be further split by approach's parameters. The figure was captured from CSVR tool developed by Probst and Reymond. \cite{Daniel2018Exploring}}} \label{CS}

\end{figure}

{
Chemical space is a concept in cheminformatics referring to the property space, which is spanned by all possible molecules adhering to a given set of construction principles and boundary conditions.\cite{kirkpatrick2004,reymond2010,oprea2002,oprea2001,reymond2015}
As shown in Fig.\ref{CS}, the chemical space considered in this work is spanned by possible molecules and their computational times, which are treated as the intrinsic properties of molecules.
Different computational schemes (e.g. hardware, software, and approaches, etc) lead to different computational times for molecule. Thus, there are various chemical spaces for a given set of molecular suits. 
Even when a specific approach is employed, there are still various chemical spaces that are generated by the parameters. For instance, different choices of DFT functionals, basis sets, quantum chemical packages etc. can generate different chemical spaces, even if the same DFT approach is employed.
}

\subsection{Cheminformatics and the employed ML models}

{
As mentioned previously that the proposed DFT chemical spaces can be further split by different computational parameters, hence, ML techniques can be employed in each split chemical space, to train the models that can be used to perform predictions. 
}
In each DFT chemical space, we employed several ML models, from simple to complex, {to benchmark their capacities and correspondingly to pick up the reliable framework} for predicting the computational times. 
There are four carefully selected ML models within this framework, they are as follows:

{
1) Random Forest (RF),\cite{breiman2001,rf1995} basing on the structural similarity. The composition proportion of a molecule respecting to the feature structures (e.g. linear, dendritic, ring, etc.) can be evaluated by the decision trees constructed by the designed feature structures via the simplified molecular input line entry specification (SMILES) codes.\cite{weininger1988smiles, weininger1989smiles, weininger1990smiles} Then the predicted computational time can be estimated by a linear combinations of computational times of feature structures. The idea behind this process is quite similar to the linear combination of atomic orbitals (LCAO) approximation in quantum chemistry.
}

{
2) Long Short-Term Memory (LSTM),\cite{Hochreiter97,gers2000} basing on the recognition of chemical formula. The connection between molecular structures and computational times is recognized by training the similar molecular suits. The molecular structures are identified via the SMILES codes using the natural language processing (NLP), \cite{morgan1991} and the bidirectional variation of LSTM model \cite{graves2005bidirectional, sundermeyer2014translation, kiperwasser2016simple} is practically implemented in this work.
}

{
3) Message Passing Neural Network (MPNN),\cite{Gilmer2017Neural} basing on the graph-based learning of spatial structures. The connection between molecular structures and computational times is recognized by training the similar molecular suits using their spatial informations. The number of basis functions, number of electrons, bond types, molecular charges etc. can be considered as a whole in this model.
}

{
4) Multilevel Graph Convolutional Neural network (MGCN),\cite{Lu2019Molecular} similar to MPNN, has advantages in the terms of generalizability and transferability.
}

{
The brief procedure is also shown in Fig.\ref{SPCSDFT}. There are more details of these four ML models in appendix section, which can be referred for the interested readers.
}


\begin{figure}[htbp]
    \vspace{0.5cm}
	\includegraphics[scale=1.025, bb=50 0 450 275]{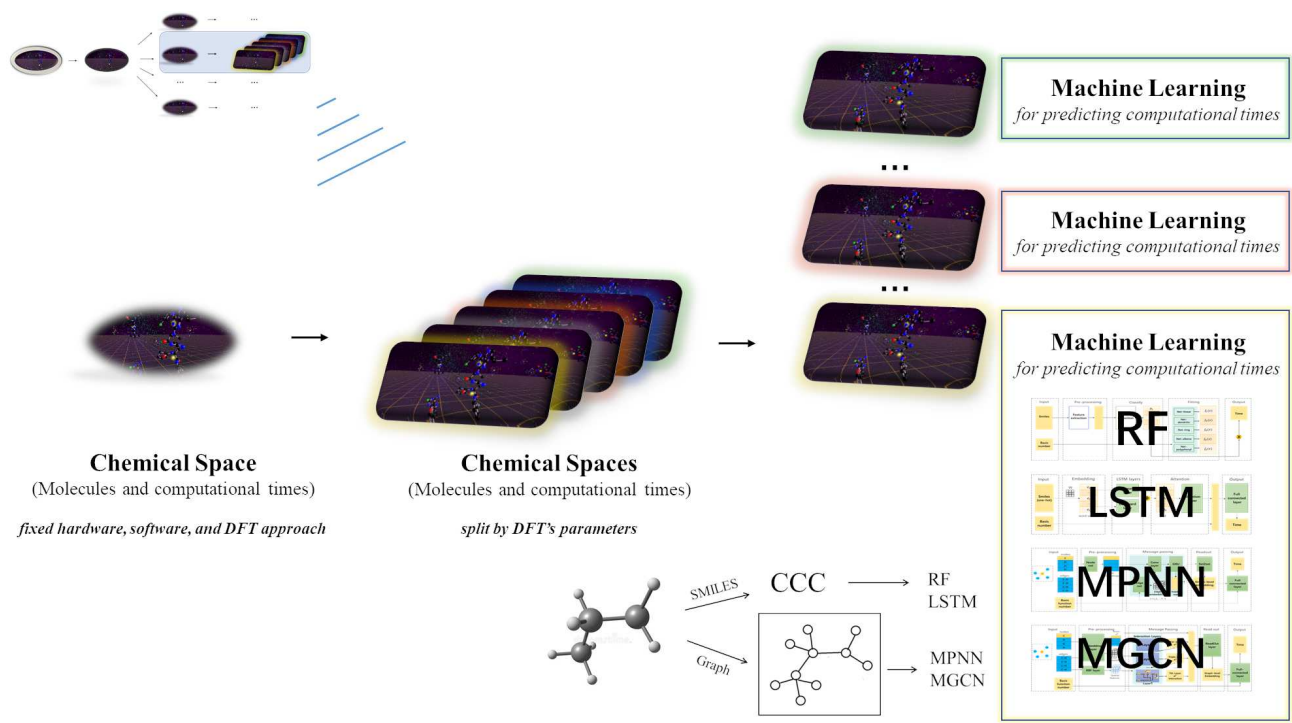}
	\caption{{Illustration of DFT chemical space, and the chemical spaces split by the computational parameters (e.g. DFT functionals, basis sets). ML techniques are employed for the purpose of predicting computational times.}} \label{SPCSDFT}
\end{figure}

\subsection{Chemical MWI ansatz and the generalization}
\begin{figure}[htbp]
	\includegraphics[scale=1.125, bb=0 0 400 275]{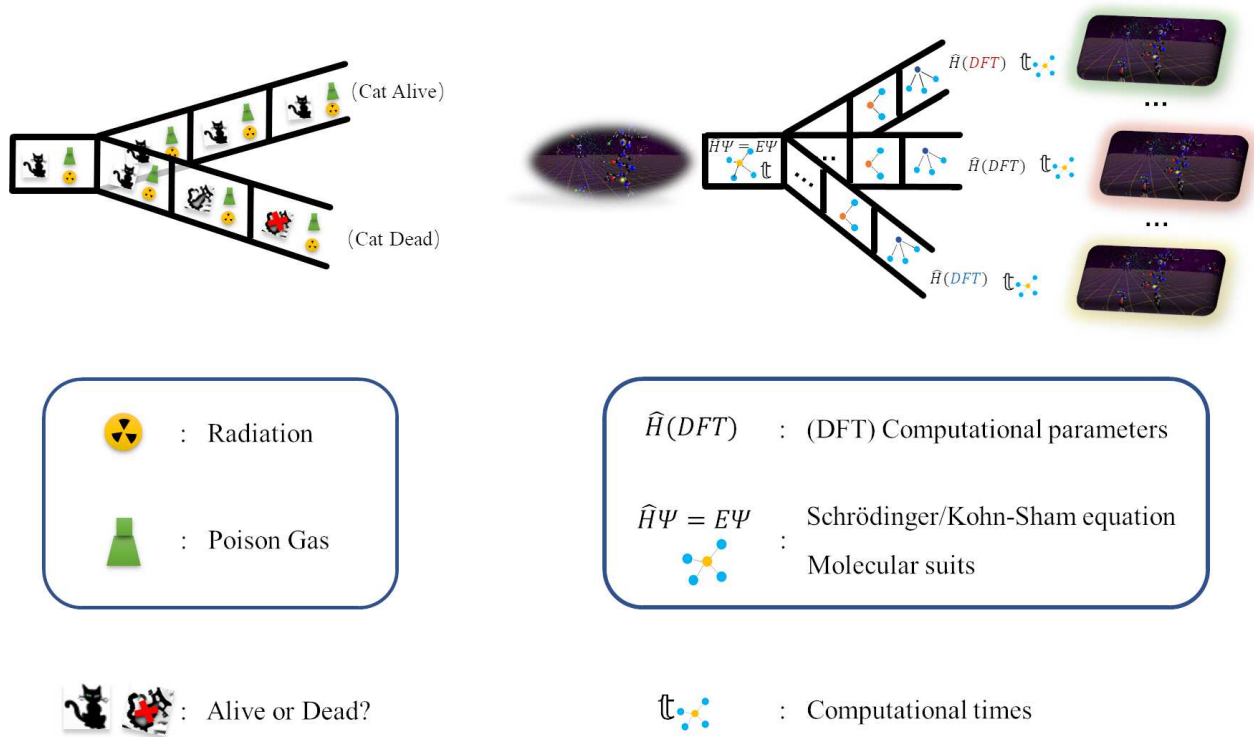}
	\caption{{Comparative illustration of original MWI using "Schr{\"o}dinger's cat" paradox \cite{schrodinger1935gegenwartige, schrodinger1935discussion} and chemical MWI ansatz used in predicting the computational times in this work. In the MWI anastz, every quantum event is a branch point; the cat is both alive and dead, even before the box is opened, but the "alive" and "dead" cats are in different branches of the universe, both of which are equally real and depended on "Radiation? $\rightarrow$ Poison Gas $\rightarrow$ Cat status" process{; there is entanglement or link between the states of cats in different spaces, i.e. if an alive cat in one space, then always a dead cat in the other space}, but do not interact with each other. In the chemical MWI anastz, different branches are split by the computational parameters that affect the actual implementation/solution of Hamiltonian operators ($\hat{H}$) following the "Parameters $\rightarrow$ KS equation \& molecule $\rightarrow$ Computational times" process. {Homologous, the  entanglement or link among chemical spaces can be also be deduced, i.e. if the computational times ($t_{ref}$) is known in one space, then the $t_{tar}$ for other space can be estimated basing on the $t_{ref}$; but still no interactions between $t_{ref}$ and $t_{tar}$ due to different spaces.}}} \label{mwi}
\end{figure}

The chemical MWI ansatz used in this paper is inspired by Hugh Everett III's MWI {multiverse ansatz}, 
\cite{RevModPhys.29.454, everett2012everett, wallace2012emergent, tappenden2000identity, bousso2012} 
in which {there is} an interpretation of quantum mechanics that asserts that the universal wave function is objectively real, and that there is no wave function collapse.
It implies that all possible outcomes of quantum measurements are physically realized in some "world" or "universe", and they all share a unique start point. 
As shown in Fig.\ref{SPCSDFT}, in chemical MWI ansatz, the \textbf{unique start point} can be the {chemical space}, in which the same molecular suits and Kohn-Sham working equation are deployed,
and every \textbf{result's world} {is the split chemical space caused by the uneven DFT parameters, which in turn result in different wave functions and different computational times as the intrinsic properties for the same molecular suits of the start point.
As illustrated in the Fig.\ref{mwi}. }

\begin{figure}[htbp]
	\includegraphics[scale=1.15, bb=25 0 400 275]{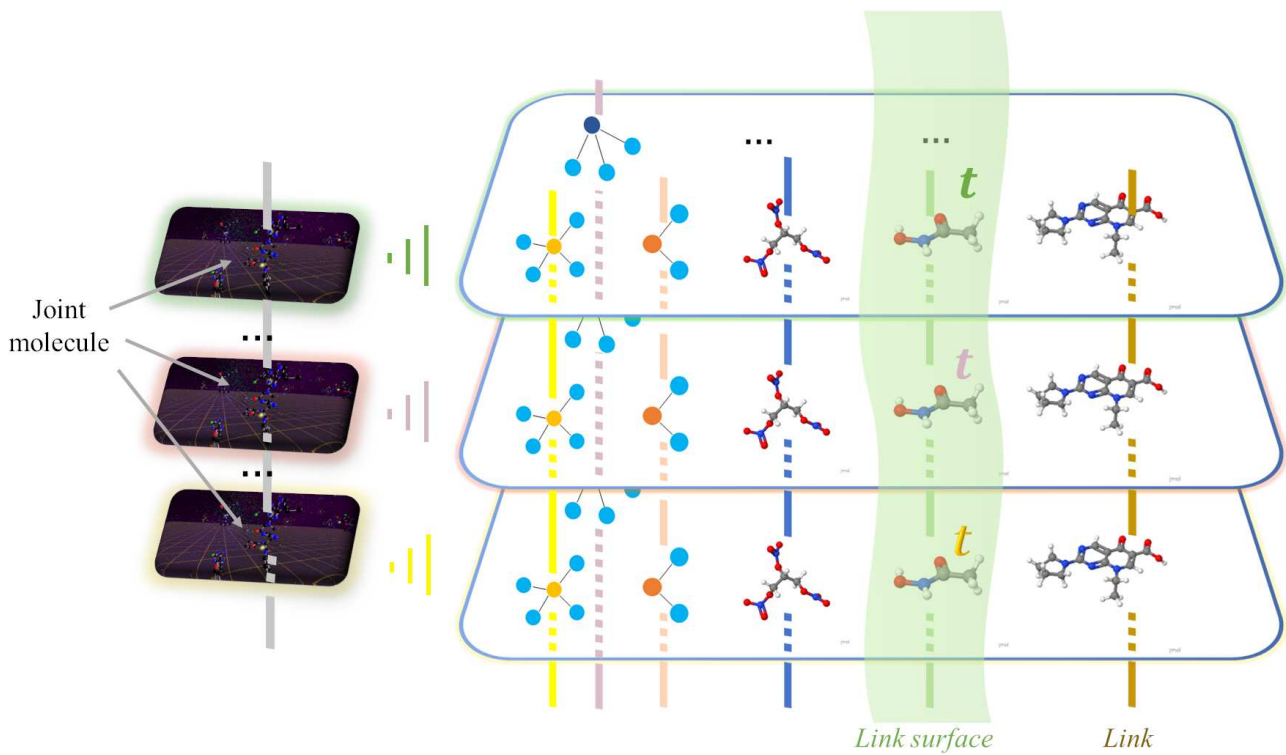}
	\vspace{-0.5cm}
	\caption{{Illustration of split chemical spaces, molecular link via joint molecule, and link surface used in the chemical MWI.}} \label{joint}
\end{figure}

{It is worthy of note that the molecular suits are included in the start point and shared by all split chemical spaces, thus, the relationships between the computational times among various split chemical spaces can be connected by the molecular suits. In the original MWI, the status of cat (i.e. observed object) in one space can always be determined/deduced by the status of cat in the other space via the so-called quantum entanglement. \cite{einstein1935can, yin2013lower} In the chemical MWI of this work, we assume the computational times (i.e. observed objects) in one space can also be deduced by the computational times in other spaces via fitting relationships. The fitting relationships can be deduced by the molecular suits or just some molecules in the suits. As shown in Fig.\ref{joint}, these molecules are defined as the joint molecules, since they represent the joints for various split chemical spaces, and can connect these split chemical spaces through the molecular link and the link surface (surface implies that it contains the computational properties like computational times).
}

{
Herein, it worth mentioning that the fitting relationships obtained via one or few molecular links may be generalized to the split chemical spaces if the solving manner for the KS equation is fixed. 
And as such, the computational times within split chemical spaces, which are caused by various combinations of basis set and DFT functional, can be well considered within this anastz.
For DFT functionals, the correction coefficients of computational times for split chemical space can be approximated by the molecular link, and the Jacob's ladder \cite{perdew2001jacob, perdew2005prescription} may be used as the entry point for further classify the functionals that did not occurred in link surface. 
In this case, the correction coefficient for the target DFT functional under the specific basis can be obtained from the expression
\begin{equation}\label{magn2}
c^{dft}=f^{dft}(tar)/f^{dft}(ref),
\end{equation}
in which the $f^{dft}(tar)$ and $f^{dft}(ref)$ are the target and reference timing data, respectively, for the molecular link. 
If the $f^{dft}(tar)$ is not available, the DFT functional within the same region of Jacob's ladder can be used as the substitution.
For basis sets, the correction coefficients can be approximated by the molecular link via a polynomial curve-fitting technique. 
In this case, the correction coefficient for the target basis set under the specific DFT functional can be obtained through the expression
\begin{equation}\label{fittedcurve}
c^{bas}=f^{bas}(x_{tar})/f^{bas}(x_{ref}),
\end{equation}
in which the $f(x)$ denotes the fitted polynomial equation using various basis sets under the specific DFT functional, and $x$ denotes the number of basis functions. If the second-order polynomial is used, then
\begin{equation}\label{fbas}
f^{bas}(x)=ax^2+bx+c
\end{equation}
can be used for calculating the timing data for the target and reference basis sets, respectively.
Nevertheless, it is worthy of note that the reliability of the correction coefficient is affected by the deviation of target basis set and reference basis set.  
Here, a similarity coefficient is introduced as a measurement of the deviation between the target and reference basis sets. 
The formula of similarity coefficient ($s$) can be expressed as
\begin{equation}\label{similarity}
s=\frac{min(N_{tar},N_{ref})}{max(N_{tar},N_{ref})}\cdot (1-|\rho_{tar}-\rho_{ref}|)\cdot J, 
\end{equation}
where $N_{tar/ref}$ represents the number of atomic basis functions, $\rho$ measures the contractions of basis sets, 
\begin{equation}
\rho = \frac{N_{pri}}{N_{con}}
\end{equation}
with $N_{pri}$ representing the number of primitive basis functions and $N_{con}$ represents the total number of contracted basis functions,
and $J$ denotes the Jaccard index \cite{DBLP:conf/icdm/MoultonJ18, levandowsky1971distance, jaccard1901etude} reflecting the similarity of orbital composition of the two basis sets. Further,
\begin{equation}
J=\frac{r}{r+p}
\end{equation}
where $r$ represents the number of identical atomic basis functions between the target and reference basis sets, $p$ represents the number of atomic basis functions in either of them.
The closer that $s$ is to 1, the closer the two basis sets are.
}


{It needs to be emphasized that the transfer learning,\cite{bozinovski2020reminder} i.e. the re-use of pre-trained ML model on new functional/basis set combinations with scaling factors, can be recognized as the rationale behind the chemical MWI ansatz. To be specific, the domain adaptation,\cite{ben2010theory} which is a subcategory of transfer learning and owns the ability to apply an algorithm trained in one or more "source domains" to a different but related "target domain", can be use to understand the Eq.\ref{magn2}-\ref{fbas} when assuming only the scaling factor is different between source domain and target domain. 
Beyond the domain adaptation, the model based transfer learning,\cite{pan2009survey} which is normally used for the transfer of neural networks on condition that the parameters can be shared between source and target domains, should also be feasible as the so-called "fine-tuning" operations.\cite{yosinski2014transferable}
Thus it can make the demand for data turns lower.
}

\subsection{Forecasting System}

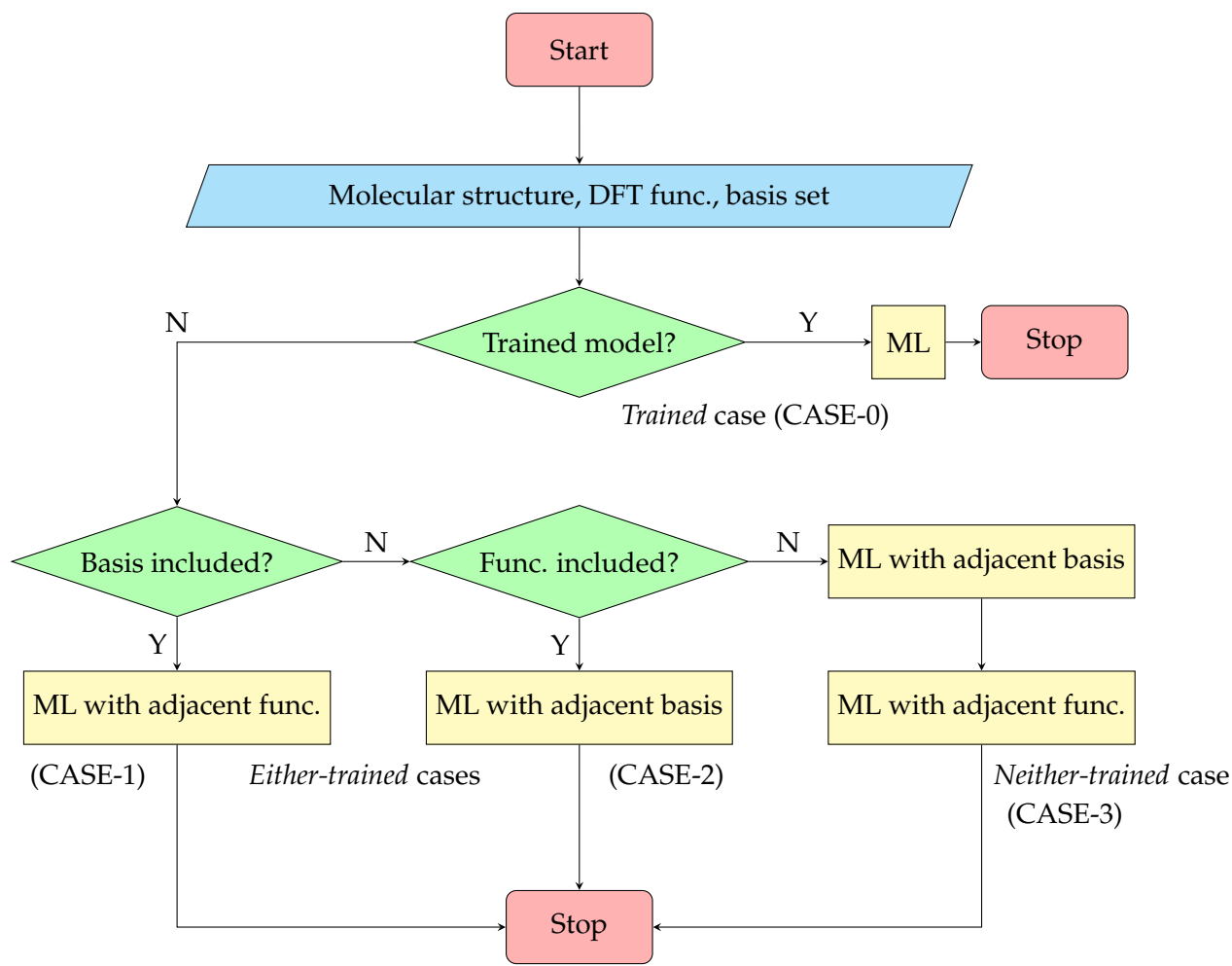
\begin{figure}
\begin{tikzpicture}[node distance=2cm]
\tikzstyle{startstop}=[rectangle, rounded corners, minimum width = 2cm, minimum height=1cm,text centered, draw = black, fill = red!30]
\tikzstyle{io}=[trapezium, trapezium left angle=70, trapezium right angle=110, minimum width=2cm, minimum height=0.85cm, text centered, draw=black, fill = cyan!30]
\tikzstyle{process}=[rectangle, minimum width=1cm, minimum height=1cm, text centered, draw=black, fill = yellow!30]
\tikzstyle{decision} = [diamond, aspect = 3, text centered, draw=black, fill = green!30]
\tikzstyle{arrow} = [->,>=stealth]
\node (start)[startstop]
{Start};
\node(in1)[io, below of = start]
{Molecular structure, DFT func., basis set};
\node(dec1)[decision,below of=in1]
{Trained model?};
\node(pro1)[process,right of=dec1,xshift=2.5cm]
{ML};
\node(dec2)[decision,below of=dec1,left of=dec1,xshift=-3.5cm,yshift=-1.0cm]
{Basis included?};
\node(pro2)[process,below of=dec2]
{ML with adjacent func.};
\node(dec3)[decision,right of=dec2,xshift=3.5cm]
{Func. included?};
\node(pro4)[process,below of=dec3]
{ML with adjacent basis};
\node(pro6)[process,right of=dec3,xshift=3.5cm]
{ML with adjacent basis};
\node(pro7)[process,below of=pro6]
{ML with adjacent func.};
\node(stop)[startstop,below of=pro4,yshift=-1cm]
{Stop};
\node(stop2)[startstop,right of=pro1]
{Stop};
\draw[arrow](start) -- (in1);
\draw[arrow](in1) -- (dec1);
\draw[arrow](dec1) -- node[above]{Y}(pro1);
\draw[arrow](pro1) -- (stop2);
\draw[arrow](dec1) -| node[above]{N}(dec2);
\draw[arrow](dec2) -- node[left]{Y}(pro2);
\draw[arrow](pro2) |- (stop);
\draw[arrow](dec2) -- node[above]{N}(dec3);
\draw[arrow](dec3) -- node[left]{Y}(pro4);
\draw[arrow](pro4) -- (stop);
\draw[arrow](dec3) -- node[above]{N}(pro6);
\draw[arrow](pro6) -- (pro7);
\draw[arrow](pro7) |- (stop);
\end{tikzpicture}
\put(-200,210) {\textit{Trained} case (CASE-0)}
\put(-345,70) {\textit{Either-trained} cases}
\put(-430,70) {(CASE-1)}
\put(-205,70) {(CASE-2)}
\put(-55,70) {\textit{Neither-trained} case}
\put(-50,55) {(CASE-3)}
\caption{Workflow of the proposed forecasting system.} \label{workflow}
\end{figure}

Fig.\ref{workflow} shows the workflow of our proposed forecasting system, which is already uploaded to Github.\cite{Fcst_sys_public}
For any input molecule, the forecasting system can have the capacity to give a predicted DFT computational time for {a specific hardware, software with any combination of DFT functional and basis set.} 
The simplest one is the trained case (denote as CASE-0), in which both the target DFT functional and the target basis set can match these of trained models. In this case, the computational time $t$ can be predicted using the matched models without any correction.
For the other cases (i.e. either-trained case and neither-trained case), the whole process can be a bit more complex, and are listed in the following:

\begin{framed}
CASE-1: Either-trained (only basis set can match).
\begin{itemize}
  \item [1)]  Identify the type (LDA, GGA, etc.) of DFT functional basing on the Jacob’s ladder.
  \item [2)]  Decide the specific reference DFT functional basing on the functional type or other preference. 
  \item [3)]  Assign the values of $f^{dft}(tar)$ and $f^{dft}(ref)$ basing on the type of DFT functional, e.g. LDA, GGA, Hybrid \& Range-separated, Meta Hybrid \& Range-separated Hybrid with values of 0.95, 1.0, 1.1, and 1.2. (see Fig.\ref{basissuit} for how these values were obtained.)
  \item [4)]  If the timing database of $DFT_{tar}/basis$ and $DFT_{ref}/basis$ calculations for link molecule were available, then the values of $f^{dft}(tar)$ and $f^{dft}(ref)$ can also obtained from the database. 
  \item [5)]  Obtain the correction coefficient $c^{dft}$ basing on eq.(\ref{magn2}).
  \item [6)]  Predict the computational time ($t_0$) of target molecule using the models of reference functional and reference the basis set.  
  \item [7)]  Correct the computational time $t$ via $ t = t_0 * c^{dft} $
\end{itemize}
\end{framed}

\begin{framed}
CASE-2: Either-trained (only DFT functional can match).
\begin{itemize}
  \item [1)]  Calculate the similarity coefficients between target basis set and reference basis sets.     
  \item [2)]  Chose the specific reference basis set with largest similarity coefficient.
  \item [3)]  Obtain the total number of AO basis ($x$) for the target molecule using the reference basis set and target basis set, respectively.
  \item [4)]  Get the two $f^{bas}(x)$ values using the $x$ equal to $x_{tar}$ and $x_{ref}$ using the fitted polynomial equation under the reference DFT functional.
  \item [5)]  Obtain the correction coefficient $c^{bas}$ basing on eq.(\ref{similarity}).
  \item [6)]  Predict the computational time ($t_0$) of target molecule using the reference basis set.
  \item [7)]  Correct the computational time $t$ via $ t = t_0 * c^{bas} $
\end{itemize}
\end{framed}

\begin{framed}
CASE-3: Neither-trained (neither DFT functional and basis set can match).
\begin{itemize}
  \item [1)]  Same as that in CASE-2.   
  \item [2)]  Same as that in CASE-2.   
  \item [3)]  Same as that in CASE-2.   
  \item [4)]  Get the two $f^{bas}(x)$ values using the $x$ equal to $x_{tar}$ and $x_{ref}$ using the fitted polynomial equation under the specific/preferred DFT functional.
  \item [5)]  Same as that in CASE-2.   
  \item [6)]  Same as 1) to 5) steps in CASE-1, then obtain the $c^{dft}$.
  \item [7)]  Predict the computational time ($t_0$) of target molecule using the reference basis set.
  \item [8)]  Correct the computational time $t$ via $ t = t_0 * c^{bas}*c^{dft} $
\end{itemize}
\end{framed}



\section{Computational details}

A self-written code, which has already been uploaded to Github \cite{Fcst_sys_public}, was used to do the training and testing calculations.
The {\sc basis set exchange},\cite{pritchard2019new} a community database for quantum chemistry electronic structure calculations was used to obtain the information of basis sets and electrons. The {\sc libxc}\cite{LEHTOLA20181}, a community database for DFT functionals was used to obtain the information of functionals.
The {\sc stk}\cite{turcani_stk:_2018} package together with the {\sc RDKit}\cite{noauthor_rdkit_nodate} package were used for generating the molecular suites. These two packages were also used for extracting and labeling properties of the molecular suites. 
The dataloader in {\sc Tencent Alchemy Tools} \cite{quantumlab} was modified to porting the possible information when training or testing models.
All the calculations were implemented by the {\sc Gaussian}\cite{g09} (version 09.D01), {\sc NWChem}\cite{apra2020nwchem} (version 7.0.0), {\sc GAMESS}\cite{GAMESS} (version 2018.R3), or {\sc OpenMolcas}\cite{fdez2019openmolcas} (version 8.4) packages.
For all the Gaussian calculations, the Sugon W760-G20 server were used with two-way Intel Xeon E5-2680V3 processors (24 cores in total) and 128GB memory. For the other calculations, the Sugon CB60-G16 server were used with two-way Intel Xeon E5-2680V2 processors (20 cores in total) and 64GB memory.
The self-written scripts using {\sc python} with {\sc numpy}, and {\sc pytorch}\cite{paszke2017automatic} were used for automatic execution of the calculations, assembling the data, and analyzing the results. 

{The molecular suit of the RF with FNN model was artificially designed, containing 108 molecules with typical structures.
When training the models, five typical molecular suits (i.e. single/double-bond linear, branch, ring, and polyphenyl) were used as the training suits for RF models.
The molecular suit of Bi-LSTM, MPNN, and MGCN models were sampled from the DrugBank dataset.\cite{wishart2006drugbank, wishart2008drugbank, law2014drugbank}.
The criteria applied to select the molecules for the training and test sets from DrugBank dataset were as following: 

a.	Concentrate the DrugBank suits : divide all molecules of DrugBank into groups basing on the rows of periodic table of elements, manually select the desired groups (e.g. groups that contains the first two rows elements) to form the molecular suits.

b.	Generate the incremental molecular subgroups : further divide the selected molecular suits into incremental sub-groups basing on the number of atoms without the H atom.

c.	Get the training and testing reservoirs : randomly select the training molecules and testing molecules in each sub-groups with fixed ratio (e.g. 4:1) to form the training reservoir and testing reservoir. 

d.	Obtain the training and testing suits : choose the appropriate number of molecules from the training/testing reservoir to form the training/testing suits that in practical usages.

When training the models, {the width of the hidden layers was set to five meaning that every hidden layer has five neurons.} The mean absolute error (MAE) loss, 
\begin{equation}
MAE=\frac{1}{N}\sum_{i=1}^{N}\lvert \hat{y}^{(i)}-y^{(i)} \rvert
\end{equation}
was used as the target function, and the mean relative error (MRE),
\begin{equation}
MRE=\frac{1}{N}\sum_{i=1}^{N}\frac{\lvert \hat{y}^{(i)}-y^{(i)} \rvert}{y^{(i)}}
\end{equation}
was used for evaluating the performance of the model. For an input sample $i$, the $\hat{y}^{(i)} $ denotes the model output and the $y^{(i)} $ denotes the real value of the prediction target. Adam optimizer\cite{kingma_adam_2014} was used in minimizing the loss.}

\section{Results and discussions}

\subsection{Predicting with pre-trained models}

\begin{figure}[htbp]
\includegraphics[scale=0.50, bb= 50 0 900 500]{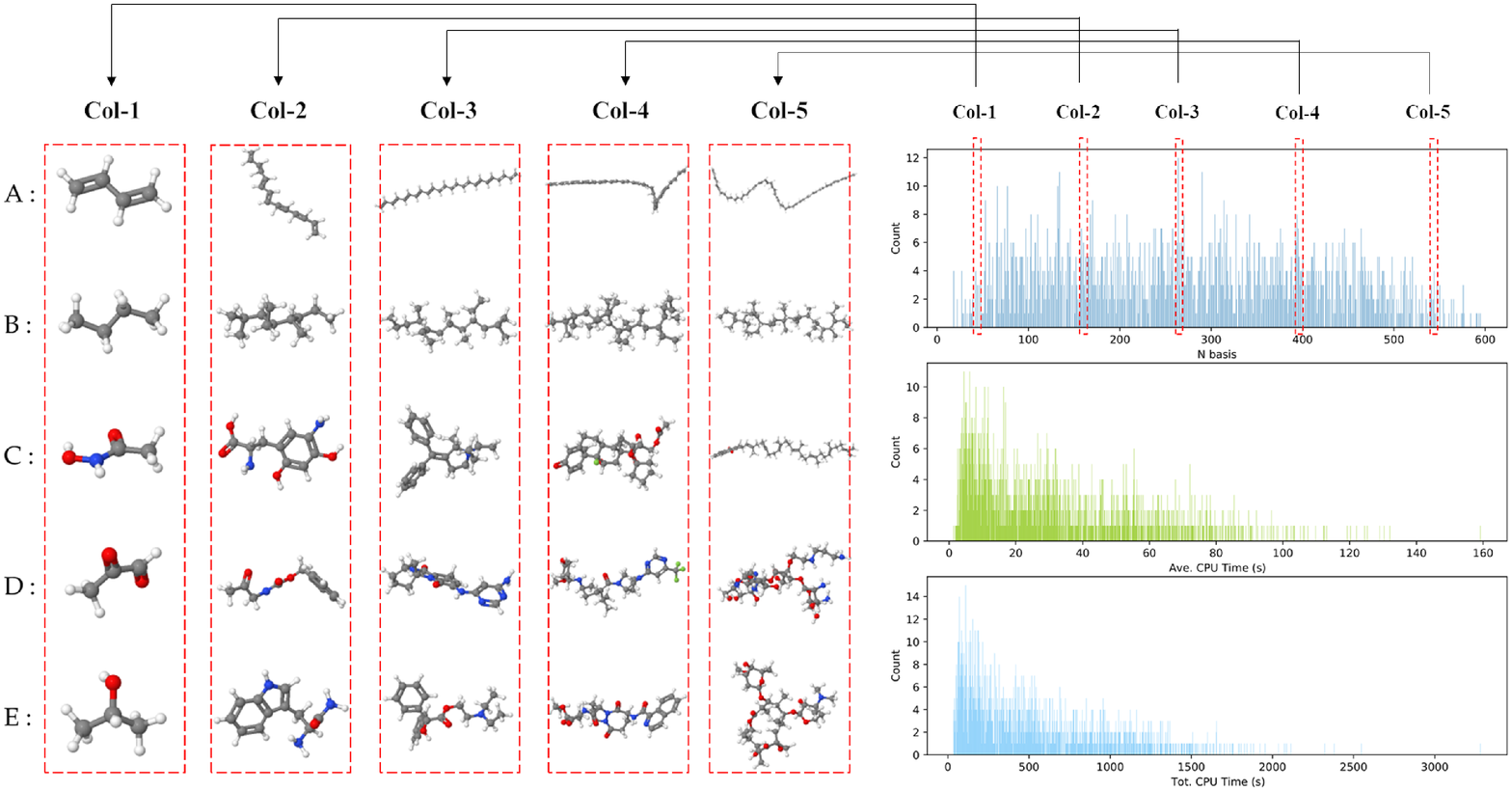}

\vspace{0.4cm}
\includegraphics[scale=0.35, bb= -370 0 2000 570]{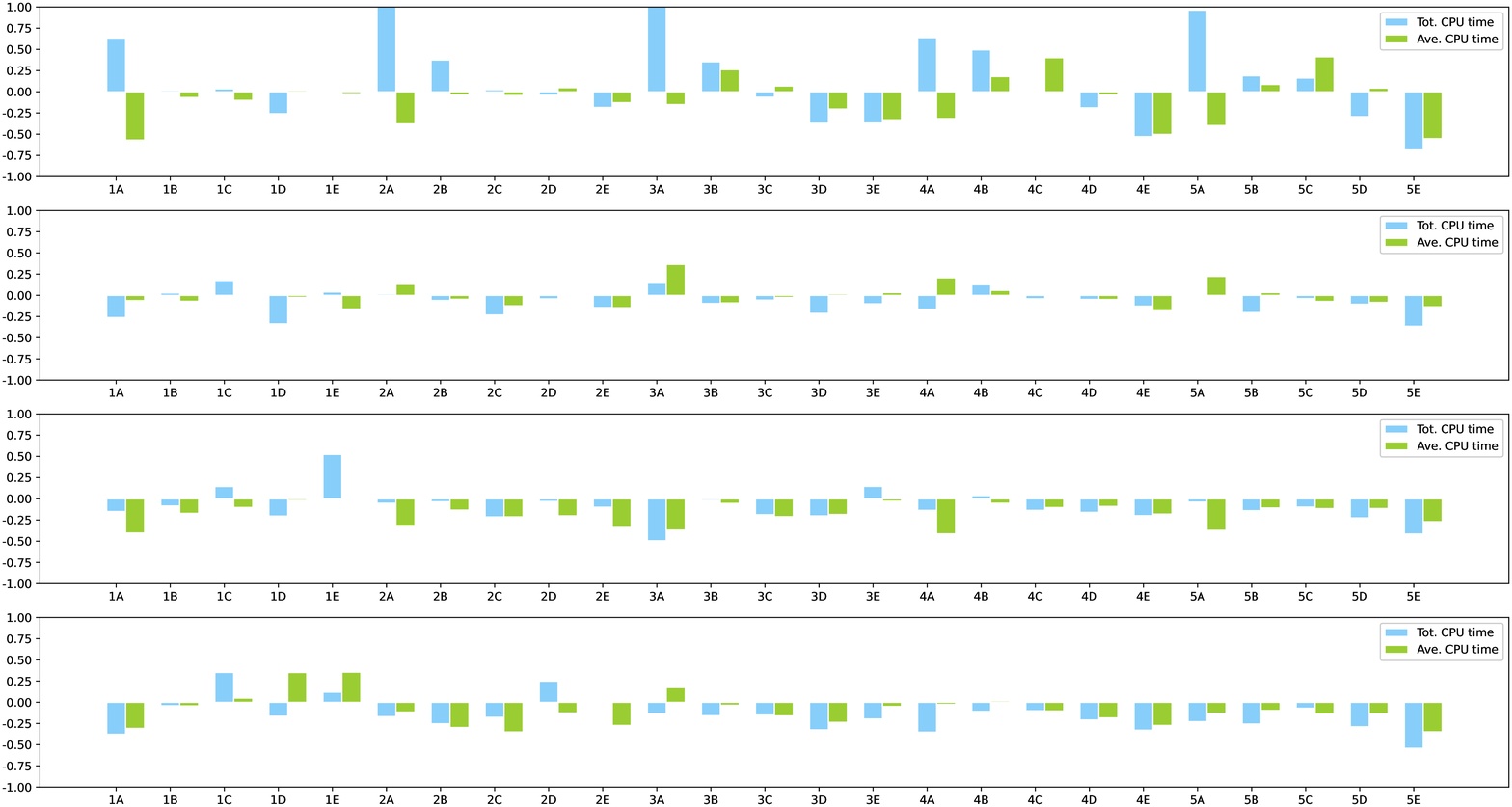}
\put(-685,232){RF   }
\put(-685,170){LSTM }
\put(-685,110) {MPNN }
\put(-685,47) {MGCN }

\vspace{-0.5cm}
\caption{{\textbf{Top left :} Illustration of sampled geometrical configurations of testing molecules. A-row and B-row are the molecules from the RF training suits, while C-, D-, and E-row are the molecules randomly selected from the {\sc DrugBank} database. The molecules in the same column possess almost the same total number of basis sets but different geometrical configurations when using the 6-31G basis sets. \textbf{Top right :} The distributions of the total number of basis functions and the CPU times for all the molecule suits, as well as the positions of the sampled molecules. \textbf{Bottom :} 
The relative errors of the predicted CPU time for all the sampled molecules. }}\label{discrimination-1}
\end{figure}

{
At the beginning, the ML models in foresting system were evaluated, to assess their capacity of distinguishing of molecules.
Several molecules, which have nearly the same total number of basis functions but with different geometrical configurations, were used for illustrations. As shown in Fig.\ref{discrimination-1}, there are 25 sample molecules that are grouped into 5 rows: A-row and B-row are the molecules from the RF training suits, and C-, D-, and E-row are the molecules randomly selected from the DrugBank database. The molecules in the same column own almost the same total number of basis functions but different geometrical configurations when using the 6-31G basis sets \cite{ditchfield1971self} with M06-2x DFT functional \cite{zhao2008m06}.
}

{
Fig.\ref{discrimination-1} presents the predicted total CPU time, CPU time averaged to each SCF iteration, and their relative errors. 
It can be seen that the predicted total CPU times using LSTM models show the best accuracy, and the calculated MRE of testing 25 molecules is about 0.13 (Tab.\ref{groupMRE}). 
The results of MPNN, MGCN models also show relatively good accuracy with MRE values were 0.17 and 0.21, respectively.
The results of RF shows the poorest results with MRE is 0.37. 
It can also be observed in Fig.\ref{discrimination-1} that a similar tendency can be observed for the averaged CPU times. 
To evaluate the capacity of the proposed method in identifying molecules with different size, the MRE values of total CPU times for column grouped molecules with different sizes were also listed in Tab.\ref{groupMRE}. 
A decreasing accuracy was observed in LSTM $\rightarrow$ MPNN $\rightarrow$ MGCN $\rightarrow$ RF order. Except the RF model, both LSTM and MPNN/MGCN models can give the reliable predictions for each column's molecules. It implied that these three models had the capacities to consider the changes of computational times that caused by the structural differences. }

\begin{table}[htbp]	
	\caption{{The MRE values of total computational times for testing molecular suits and grouped molecules (shown in) with different number of basis functions.}} 
	\label{groupMRE}	
	
	\begin{tabular}{cccccccccccccccccccccccc}
		\hline
		\hline
\multirow{2}*{Model}  &   & Testing suit  &  & \multirow{2}*{Col-1} & & \multirow{2}*{Col-2}  & & \multirow{2}*{Col-3} & & \multirow{2}*{Col-4} & & \multirow{2}*{Col-5}  \\	 
                      &    & (25 samples)&  &       & &        & &       & &       & &        \\
		\hline   
{RF}   &  & 0.37 &  &  0.89 &  & 0.29   &  &  0.06  &  &  0.23&  &  0.36     \\
 		\cline{1-2}     	
{LSTM} &   & 0.13 &  &  0.12  &  & 0.10   &  &  0.11  &  &  0.15   &  &  0.15     \\
 		\cline{1-2}              
{MPNN} &   & 0.17 &  &  0.17  &  & 0.06   &  &  0.16  &  &  0.16   &  &  0.27     \\
 		\cline{1-2}                      
{MGCN} &  & 0.21 &  &  0.25  &  & 0.16  &  &  0.17  &  &  0.25  &  &  0.24   \\                                   
        \hline
	\end{tabular}
\end{table}

\begin{figure}[htbp]
\includegraphics[scale=0.60, bb= 60 0 740 350]{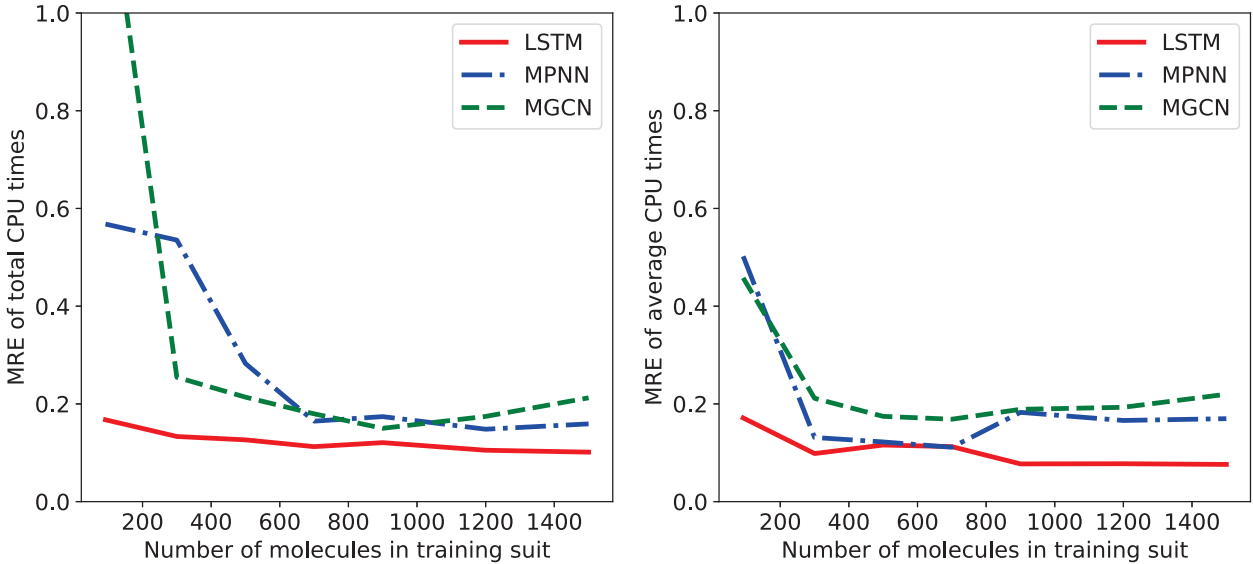}
\vspace{-0.5cm}
\caption{Illustrations of the MRE of testing molecules for total/average CPU times along with the the total number of molecules in the training suits. }\label{magnitude}
\end{figure}

{
After evaluating the classification capacities of the models in the proposed foresting system, we further checked how the magnitudes of the training suits affect the accuracy. Because the training suits in the RF models were fixed and artificially designed, the other three models (i.e. LSTM, MPNN and MGCN) were checked in this part by increasing or decreasing the molecules in the training suits. Herein, the starting point of the training suits only includes four typical molecular suits that were used in the RF models, then the molecules from the {\sc DrugBank} database were gradually added. 
The testing suit contained 116 molecules that extracted from the DrugBank database.
The results were shown in Fig.\ref{magnitude}. It can be seen that the MRE values of testing suits can already attain lower than 0.2 value for all these three ML models with less than 1000 molecules in the training suits. This magnitude of training suits was far less than that the one obtained in image recognition, because the SMILES or one-hot representations of chemical elements can already recognize the constituent of molecules. Thus only the geometric constructions need to be learned when training the models.
}

{
Assuming that one has the typical aiming molecules, e.g., PE molecules or drug molecules, then there may be ways for optimizing the training of the ML models using a fewer number of training suits, or to achieve higher accuracies.  
Here, we propose "space-specific" (S.-S.) and "space-averaged" (S.-A.) ways for training the models. In the former case, only the molecules of the same type as the aiming molecules were selected to the training suits. However, in the latter case, several types of molecules were used to ensure better generalization ability. For demonstration purposes, molecules of the PE suits (as part of molecules shown in row-A of Fig.\ref{discrimination-1}), branch suits (as part of molecules shown in row-B of Fig.\ref{discrimination-1}) and the DrugBank suits (as part of molecules shown in row-C, D, E of Fig.\ref{discrimination-1}) were used as the training and testing suits. The results for the LSTM, MPNN, and MGCN models were shown in Tab.\ref{rf_spsa}. It can be seen that the MREs of LSTM models for both S.-S. and S.-A. cases were quite close to each other, the derivations were normally less than 0.04 for both total and averaged CPU times. For the graph-based MPNN and MGCN models, the MREs of S.-S. case were in general larger than those of the S.-A. case in the small case, e.g., total CPU results of MPNN's branch (MRE was 0.63) and DrugBank/M100 (M100 denotes 100 training molecules, with MRE was 0.42,) and MGCN's DrugBank/M100 (MRE was 0.31). However, the MREs can be reduced by simply adding the number of training samples, e.g., in the total CPU results for the MPNN/MGCN's DrugBank/M300-700 cases, the MREs were reduced to around 0.15. This was line with the tendency shown in Fig.\ref{magnitude} that a certain amount of samples were needed for graph-based models. Overall, there was no obvious difference between S.-S. and S.-A. strategy when enough training samples were used (e.g. M$>$300), and a decreasing accuracy was observed in LSTM $\rightarrow$ MGCN $\rightarrow$ MPNN $\rightarrow$ RF order.
}

\begin{table}[htb]	
	\caption{The MREs of predicted total CPU times and predicted averaged CPU times for each iteration (in bracket) of the ML models using space-specific and space-averaged training approaches. } 
	\label{rf_spsa}	
	
	\begin{tabular}{cccccccccccccccccccccccc}
		\hline
		\hline
\multirow{2}*{Model} &  & Training &  \multirow{2}*{PE} & &  \multirow{2}*{Branch}  & & \multicolumn{9}{c}{Drug}    \\	 
   \cline{8-15}	
		             &  & manner   &                    & &                         & &      (M=100)      &  &     (M=300)     & &     (M=500)  & &     (M=700)        \\             
		\hline   
\multirow{2}*{LSTM}	& &	S.-S.  &  0.13 (0.05) &  & 0.11 (0.09)  &  &  0.12 (0.13) &  &  0.09 (0.12)  &  &  0.10 (0.12) & &  0.10 (0.13)  &   \\
                    & & S.-A.  &  0.15 (0.07) &  & 0.06 (0.05)  &  &  0.13 (0.13) &  &  0.11 (0.11)  &  &  0.14 (0.19) & &  0.11 (0.10)  &   \\
 		\cline{1-3}              
\multirow{2}*{MPNN}	& &	S.-S.  &  0.17 (0.11) &  & 0.63 (0.12)  &  &  0.42 (0.19) &  &  0.16 (0.15)  &  &  0.15 (0.15) & &  0.13 (0.14)  &    \\
                    & & S.-A.  &  0.21 (0.27) &  & 0.20 (0.12)  &  &  0.21 (0.19) &  &  0.16 (0.13)  &  &  0.16 (0.17) & &  0.20 (0.13)  &    \\
 		\cline{1-3}                      
\multirow{2}*{MGCN}	& &	S.-S.  &  0.14 (0.06) &  & 0.11 (0.09)  &  &  0.31 (0.12) &  &  0.20 (0.10)  &  &  0.12 (0.10) & &  0.15 (0.11)  &    \\
                    & & S.-A.  &  0.18 (0.09) &  & 0.14 (0.16)  &  &  0.16 (0.10) &  &  0.15 (0.11)  &  &  0.12 (0.15) & &  0.14 (0.15)  & &    \\                                          
        \hline
	\end{tabular}
\end{table}

\begin{figure}[htbp]
\includegraphics[scale=0.65, bb= 100 0 500 700]{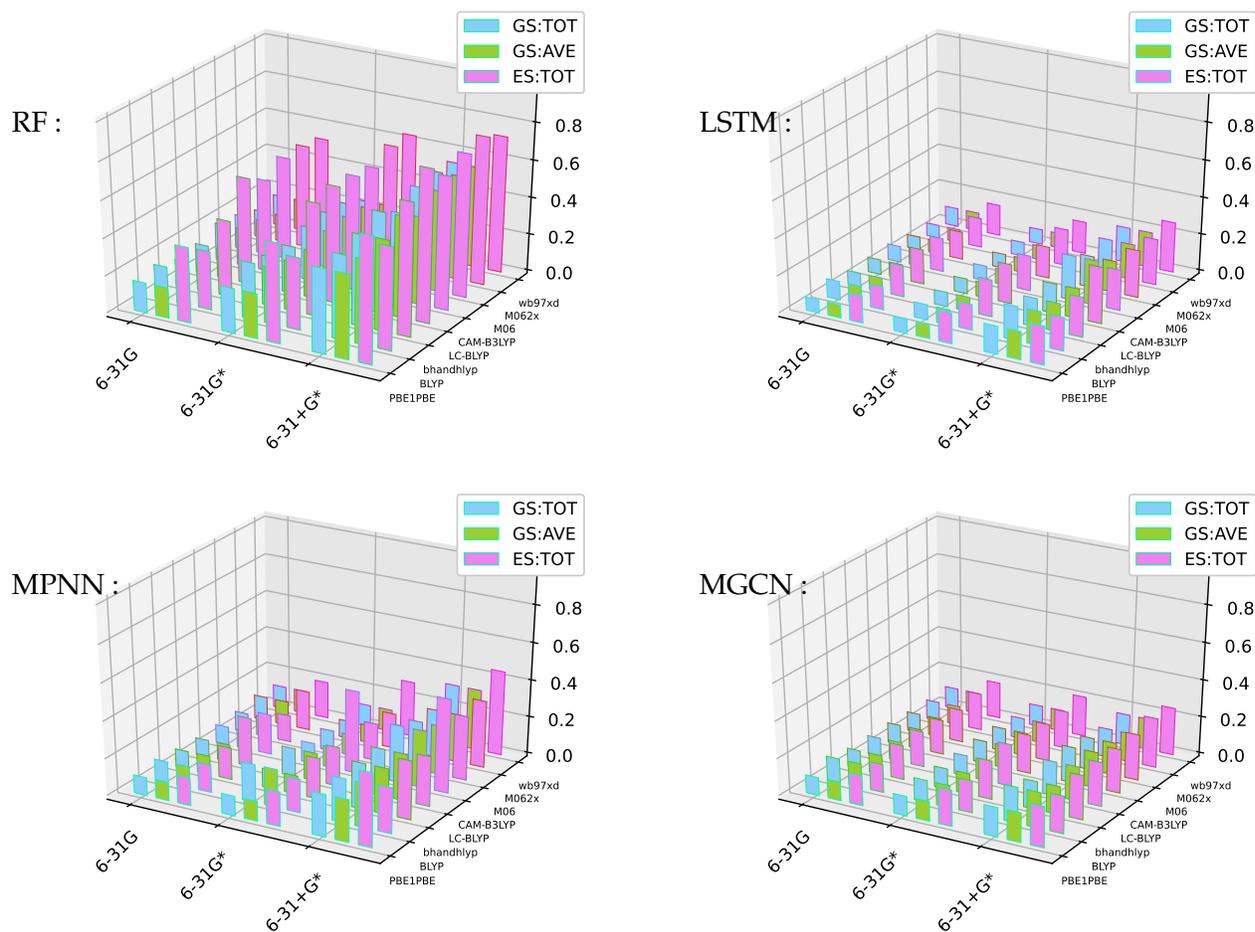}
\put(-360,375) {RF   :}
\put(-100,375) {LSTM :}
\put(-360,200) {MPNN :}
\put(-100,200) {MGCN :}
\vspace{-1.5cm}
\caption{MREs of the predicted total times of DFT calculations (cyan bars), predicted average times for each DFT iterations (green bars) and predicted total times of TDDFT calculations (violet bars) for the ML models.} \label{DFT-TDDFT}
\end{figure}

\begin{figure}[htbp]
\includegraphics[scale=0.35, bb= -400 0 1400 650]{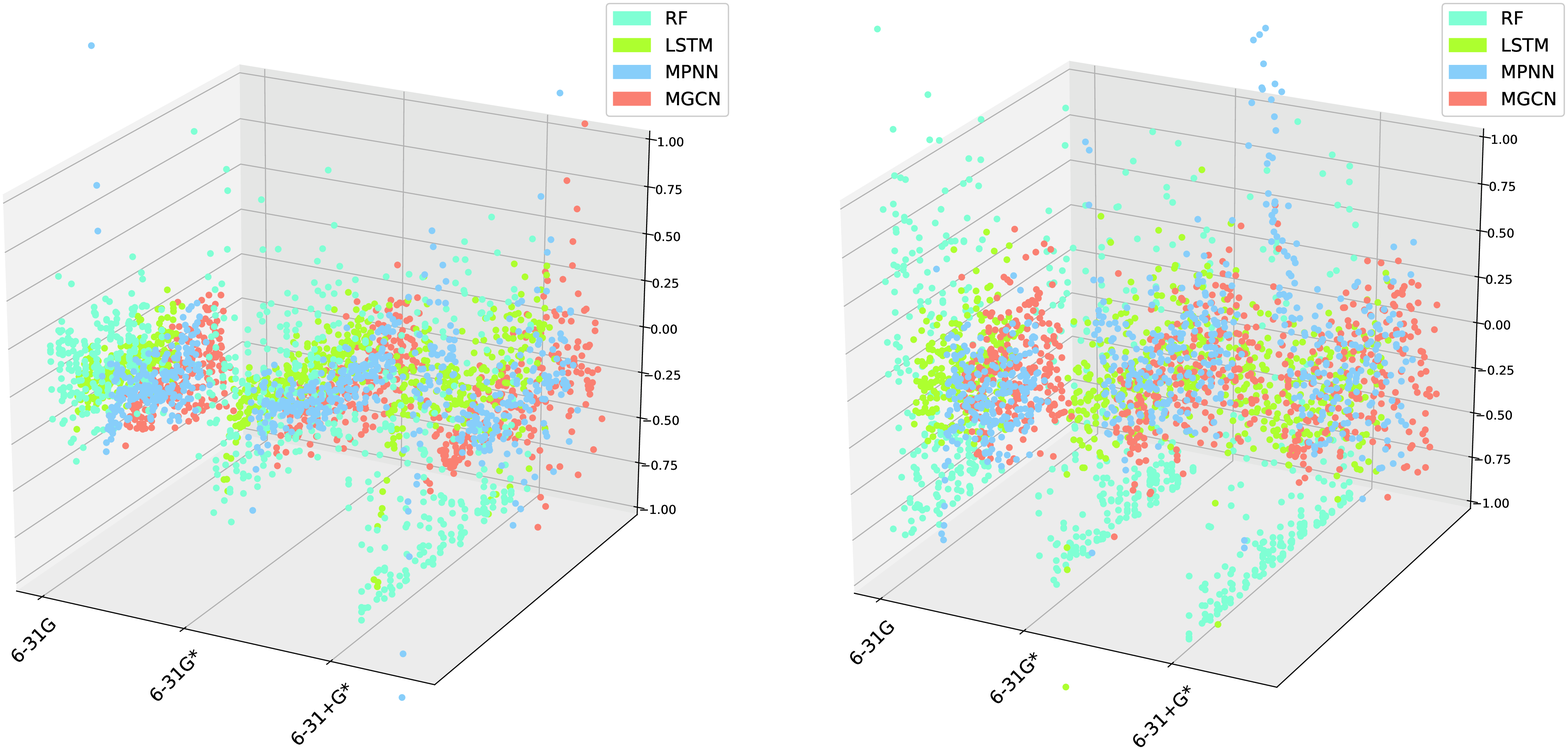}
\vspace{-1.0cm}
\caption{The MREs of the predicted total times of DFT calculations (cyan bars), predicted average times for each DFT iterations (green bars), and the predicted total times of TDDFT calculations (violet bars) for the ML models.} \label{DFT-TDDFT-ERR}
\end{figure}

{
To check the predictions for various DFT/TDDFT calculations, different combinations of popular DFT functionals ("PBE"\cite{perdew1996generalized}, "BLYP"\cite{becke1988density, lee1988development}, "bhandhlyp"\cite{becke1993new}, "B3LYP"\cite{raghavachari2000perspective}, "LC-BLYP"\cite{iikura2001long}, "CAM-B3LYP"\cite{yanai2004new}, "M06"\cite{zhao2008m06}, "M062x"\cite{zhao2008m06}, "$\omega$B97XD"\cite{chai2008long}) and Pople's basis sets ("6-31G", "6-31G*", "6-31+G*") \cite{ditchfield1971self} were used for this purpose. The results for the total/average CPU times of ground state DFT calculations, and total CPU times of singlet excited state TDDFT calculations, respectively, were illustrated in Fig.\ref{DFT-TDDFT}. 
The S.-S training approach was used in these calculations with 1000 training molecules and the selected 116 testing molecules. 
Overall, it can be concluded that MGCN $\sim$ LSTM $>$ MPNN $>$ RF in terms of overall performance. To be specific, in the case of predicting total CPU times of ground state DFT calculations (Fig.\ref{DFT-TDDFT}), all the four models showed good predicting capacities when using 6-31G basis sets with MREs around 0.10, while the MREs increased when polarization and diffused functions were added. For instance, the average MREs of RF model even turned to about 0.51 for the 6-31+G* case, and 0.30 for the MPNN case. Meanwhile, the MGCN and LSTM can still guarantee reliable predictions with MREs around 0.17. The results of MGCN showed better stability than those of the LSTM, it may have benefited from the graph-based learning approach of MGCN, in which the nature of its convergence manner (e.g. total iterations) can also be implicitly integrated in learning of the total CPU times. 
A similar tendency (MGCN $\sim$ LSTM $>$ MPNN $>$ RF) was observed for the prediction of total CPU times of singlet excited state TDDFT calculations with lightly augmented MRE values for all these models. 
Apart from the MREs for the predictions, the scattered error distributions were also illustrated in Fig.\ref{DFT-TDDFT-ERR} for the M06-2x functional with these three basis sets.
It can be seen that the MGCN and LSTM show good stability when most scatters were located in the $\pm 25\%$ regions, regardless of the predictions of ground or excited states.
} 
 
\begin{figure}[htbp]
\includegraphics[scale=0.525, bb=-50 0 700 700]{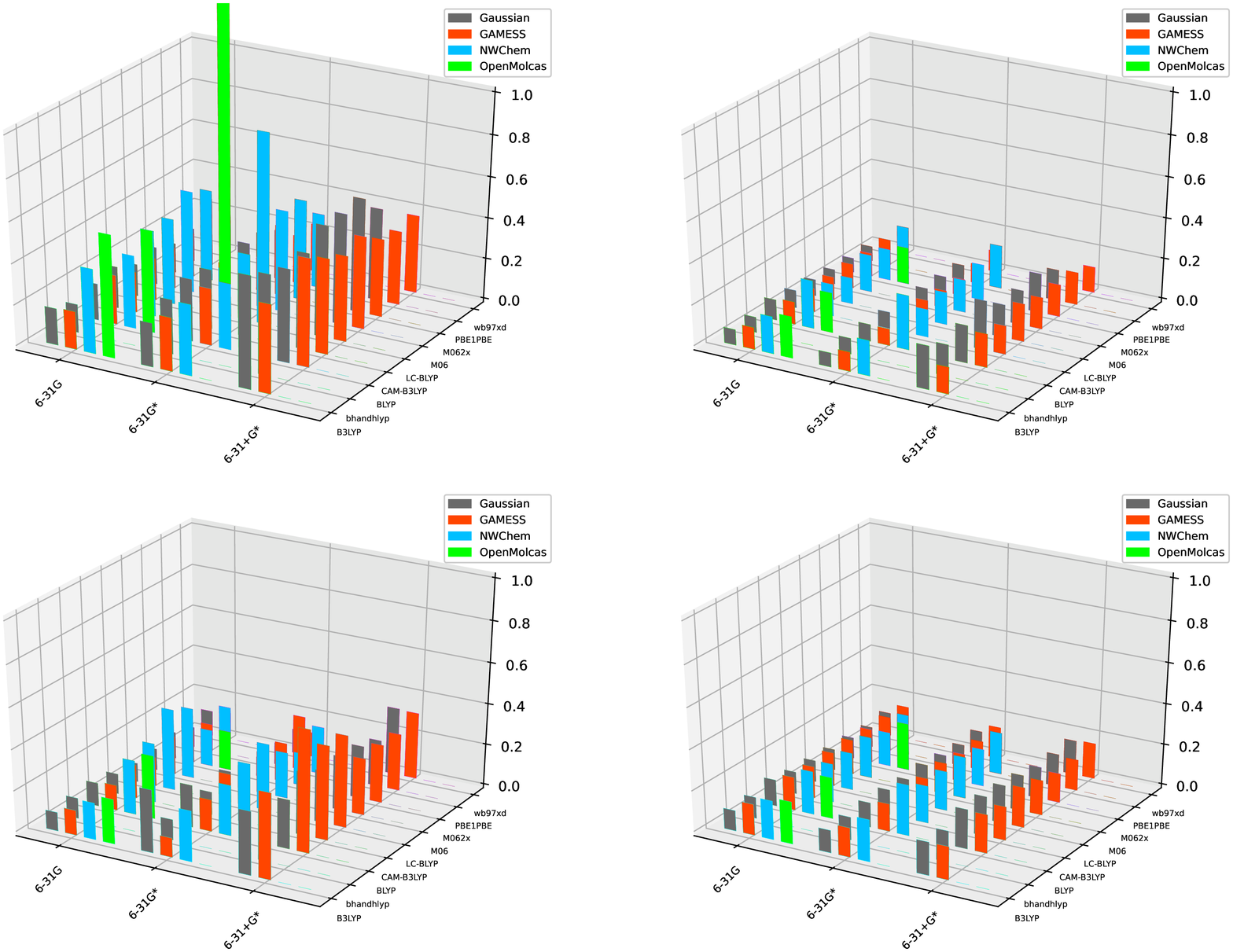}
\put(-430,350) {RF:}
\put(-170,350) {LSTM:}
\put(-430,152) {MPNN:}
\put(-170,152) {MGCN:}
\vspace{-0.7cm}
\caption{Illustrations of the MREs of the predicted total times of DFT calculations using different QC packages.}\label{MREsQCs}
\end{figure}

Before the end of this subsection, the NWChem, GAMESS, and MOLCAS packages were interfaced with the forecasting system using the same four models, to check the predicting capacity for the general quantum chemical package. The calculated MRE results are shown in Fig.\ref{MREsQCs}. It can be noticed that all the benchmarked packages basically resemble with each other when using the same ML models. The LSTM and MGCN models still showed the best precision for predictions, and thus, can be deployed in the forecasting system as the working models.
Nevertheless, we should mention that the DFT calculation time is assumed as the intrinsic property for the molecule, thus the single strong learners like MGCN/MPNN performs better than RF for this specific feature. The RF model is more suitable to tackle the noisy data, where the unknown or unobtainable features play roles. The LSTM shows good predictive ability, because the recognized ordered words already connects the structure via SMILES code thus it can be affiliated as single strong learner similar as MGCN/MPNN.

\subsection{Predicting without pre-trained models}
 
\begin{sidewaysfigure}
\includegraphics[scale=0.70, bb=-10 0 1000 500]{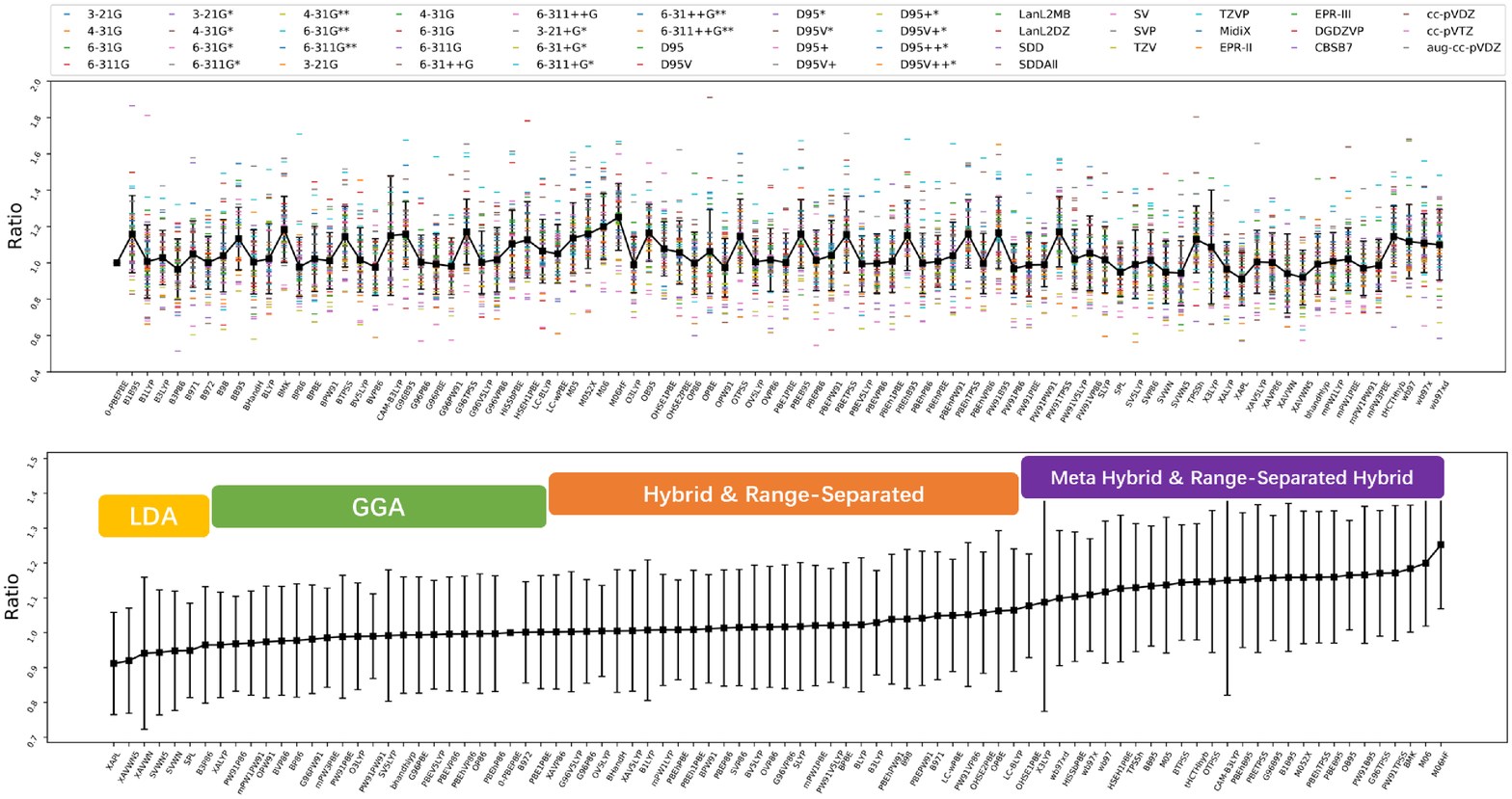}
\vspace{-0.7cm}
\caption{\textbf{Top:} Illustrations of the ratios (colored bar) between CPU times of various DFT functionals and those of the PBE functionals for listed basis sets as shown in the legend box. \textbf{Bottom:} The ratios can be re-ordered and can be roughly partitioned for different types of DFT functionals referred to as the Jacob ladder).}\label{basissuit}
\end{sidewaysfigure}

\begin{sidewaysfigure}
\includegraphics[scale=0.73, bb=0 0 1200 400]{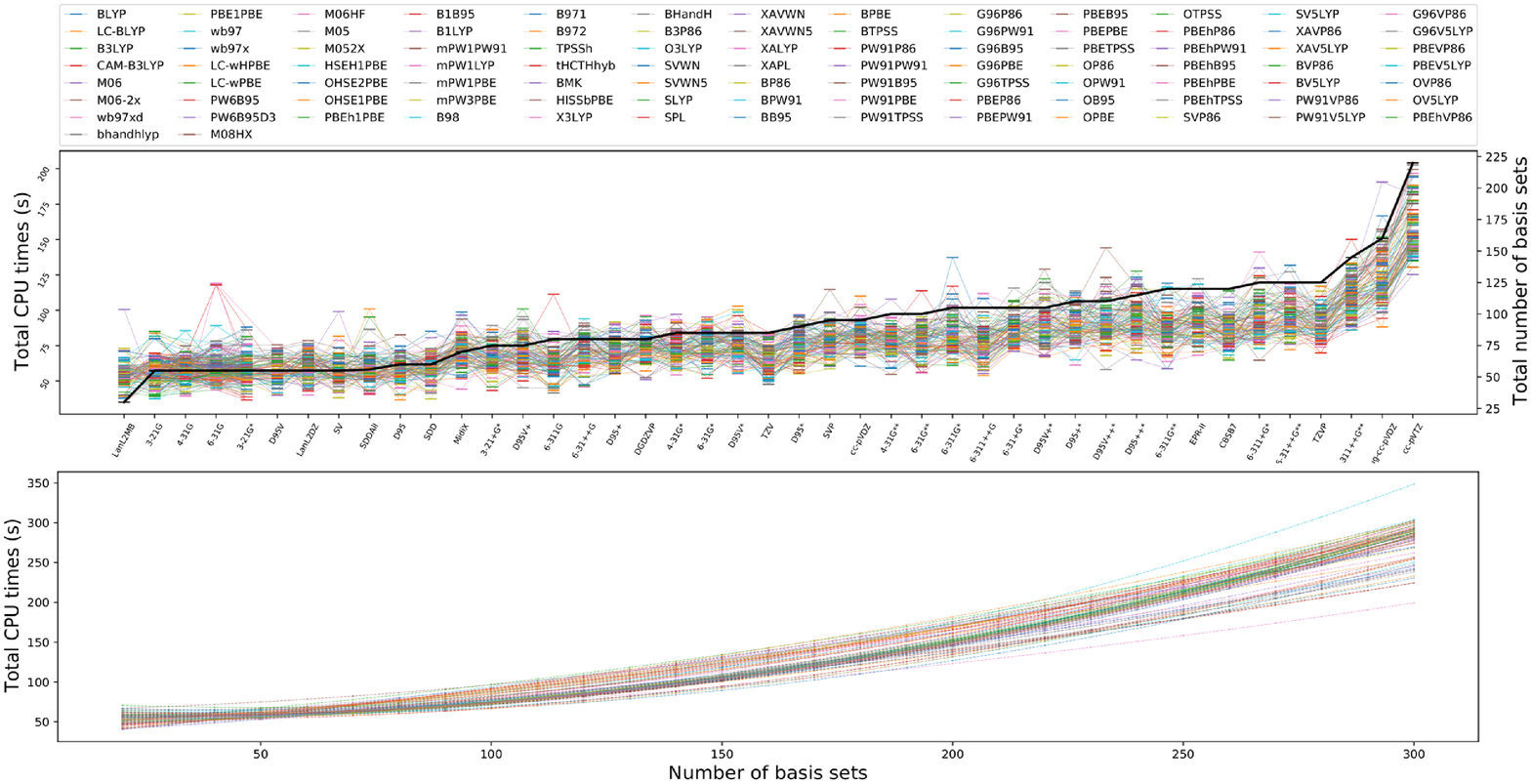}
\vspace{-0.7cm}
\caption{\textbf{Top:} Illustrations of relationships of computational CPU times and different basis sets for listed DFT functionals and basis sets. \textbf{Bottom:} Illustrated poly-fitted curves between computational CPU times and number of basis sets for every listed DFT functionals.}\label{dftsuit}
\end{sidewaysfigure}

{All the aforementioned ML models were trained in the given chemical space, now we continue to show how the forecasting system give the predictions without using pre-trained models (i.e., either-trained or neither-trained models) of the given chemical space using the chemical WMI anastz. 
As we mentioned in section II-C, a plethora of different combinations of DFT functional/basis set could be the main obstacle for general forecasting.
Thus, it is necessary to benchmark how the DFT functionals and basis sets play their roles in the practical calculations, separately. 
Herein, the relation between CPU times and parameters of DFT functionals or basis sets were shown in Fig.\ref{basissuit} and Fig.\ref{dftsuit}, respectively, via the sampled link molecule (i.e. simple CH$_3$CONHOH as shown in Fig.\ref{joint}). 
}

{
Fig.\ref{basissuit} shows the ratios between CPU times of various functionals and those of PBE functional were evaluated for the DFT functionals. It can be seen that the variations of DFT functionals did not change the CPU times a lot, e.g. the average magnifications of CPU time were in the range of 0.90 to 1.2. Furthermore, the increase of various DFT functionals can be roughly assigned to four regions referred to as the Jacob ladder. For instance, the LDA, GGA, Hybrid \& Range-separated, Meta Hybrid \& Range-separated Hybrid regions with magnifications of about 0.95, 1.0, 1.1, and 1.2, respectively. }

{For the basis sets, it can be observed in Fig.\ref{dftsuit} (top) that the computational CPU times (colored bars) matched well with the dimension of basis sets (solid line). Thus, in accordance with the change of basis sets, general polynomial curve-fitting techniques (Fig.\ref{dftsuit}, bottom) can be employed to roughly consider the change of computational times.}

\begin{figure}[htbp]
\includegraphics[scale=0.475, bb=-200 0 900 700]{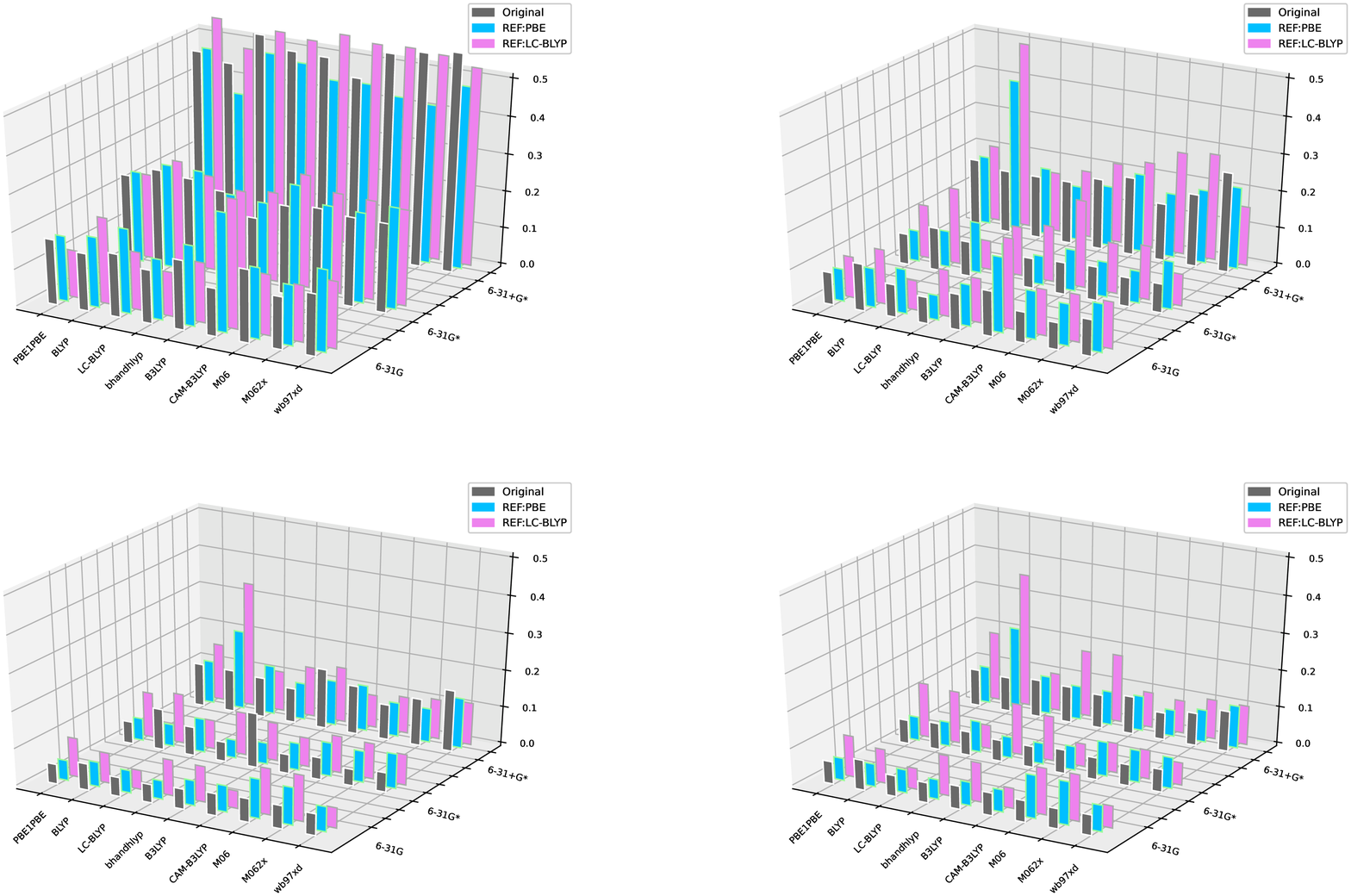}
\put(-520,330) {RF   :}
\put(-250,330) {LSTM :}
\put(-520,155) {MPNN :}
\put(-250,155) {MGCN :}
\vspace{-0.5cm}
\caption{{Illustrations of the MRE of predicted total CPU times with trained models (original) and either-trained models (REF:PBE and REF:LC-BLYP), respectively.}}\label{ref_notrained1}
\end{figure}

\begin{figure}[htbp]
\includegraphics[scale=0.37, bb=410 0 900 270]{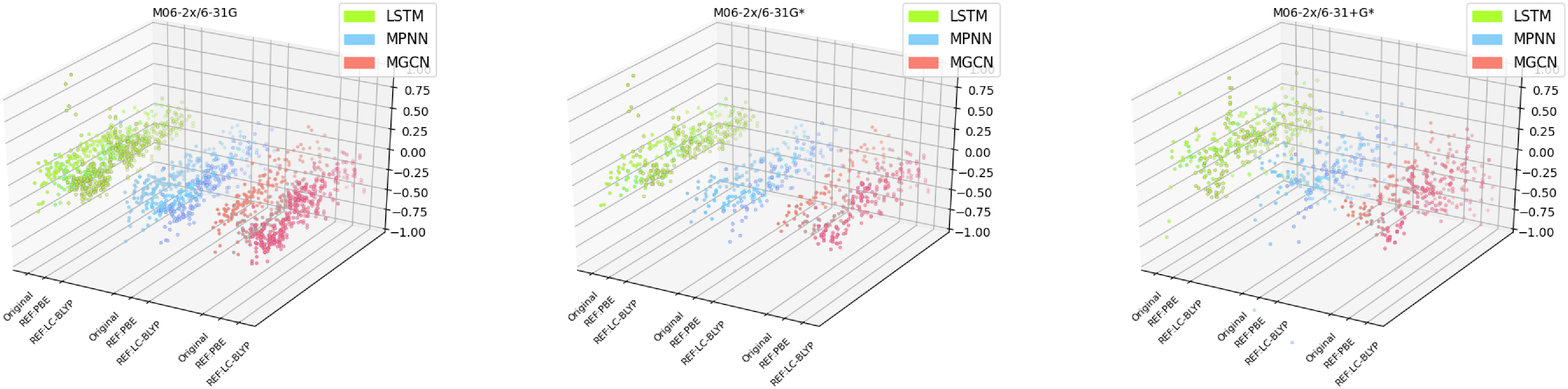}

\includegraphics[scale=0.38, bb=400 0 900 190]{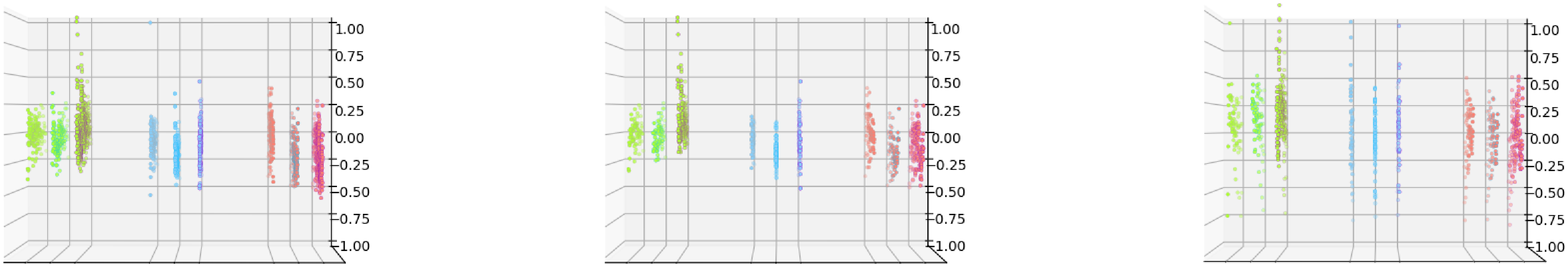}
\vspace{-0.5cm}
\caption{{Illustrations of the error scatters of predicted total CPU times using LSTM, MPNN, and MGCN models for the M06-2x cases in Fig.\ref{ref_notrained1}. Both the stereoscopic (top) the radial (bottom) distributions are shown.}}\label{ref_notrained1-ERR}
\end{figure}

{
For the case of either-trained predictions with the trained ("trained" means the parameter was involved in pre-trained models) basis sets and untrained ("untrained" implies that the parameter was not involved in pre-trained models) DFT functionals, the trained models can be used to give the predictions with magnifications based on the types of DFT functionals.
For instance, there were nine functionals together with three basis sets in Fig.\ref{DFT-TDDFT}. Suppose only the models of DFT functionals of PBE and LC-BLYP were trained, then all the timing predictions for the remaining DFT functional/basis set combinations can be deduced using these existing models as the references. 
The obtained results were shown in Fig.\ref{ref_notrained1}. 
It can be found that most derivations of predicted times were only slightly larger than the original ones except in the BLYP case, in which convergence behavior was quite different from other models. 
The MREs of predicted times with REF:PBE (PBE functional as the reference) and with REF:LC-BLYP (LC-BLYP functional as the reference) can still be less than 0.2 for LSTM and MPNN/MGCN models. The scatter diagrams of relative errors were also illustrated in Fig.\ref{ref_notrained1-ERR} for the M06-2x functional case as an instance. 
It can be seen that the error distributions of REF:PBE and REF:LC-BLYP cases were quite similar to the original ones for all the three basis sets. The MGCN showed the best accuracy among these three ML models with relative errors mostly distributed within +25\% to -25\% regions for the original case, and there were only 5\% to 20\% displacement for the REF:PBE and REF:LC-BLYP cases. 
}

\begin{figure}[htbp]
\includegraphics[scale=0.8, bb=280 0 450 375]{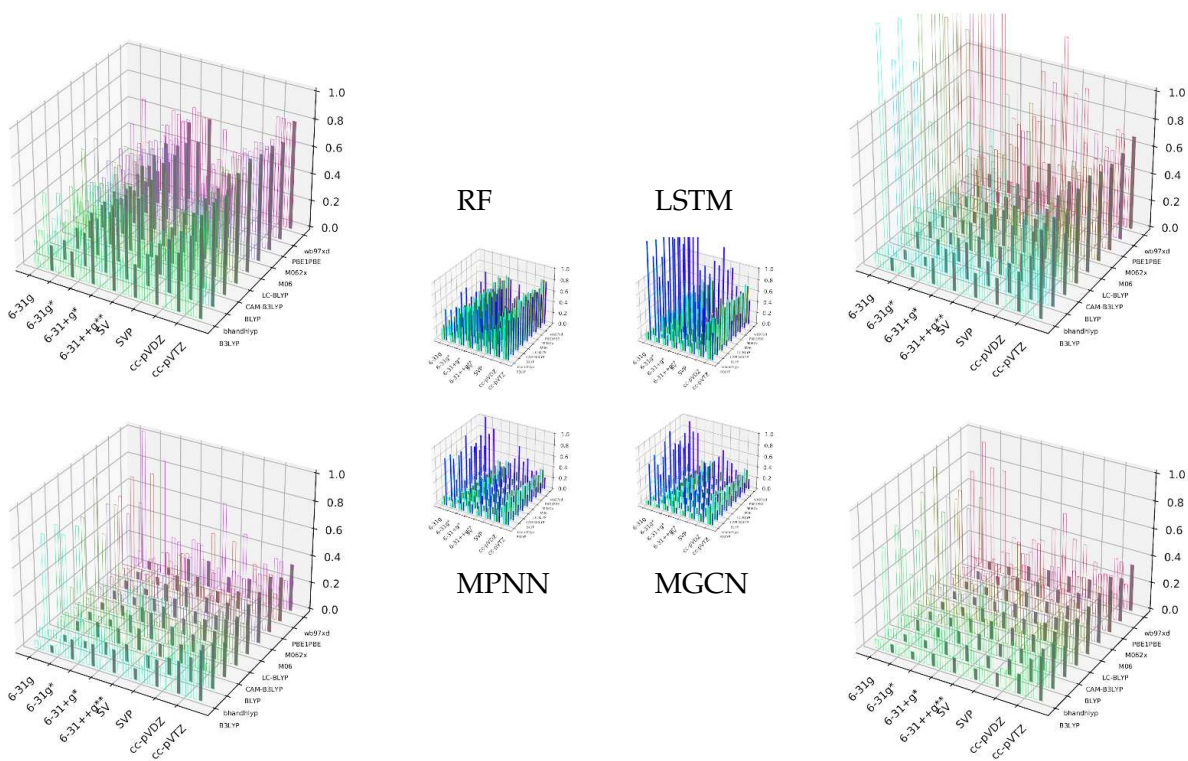}
\put(-125,225) {RF}
\put(-50,225) {LSTM}
\put(-125,80) {MPNN}
\put(-50,80) {MGCN}
\vspace{-1.0cm}
\caption{{Illustrations of the MREs of predicted total CPU times using RF, LSTM, MPNN and MGCN models with 6-31G, 6-31G*, 6-31+G* reference basis sets (center), and the similarity-corrected ones for practical predictions (outside).}} \label{ref_notrained2}
\end{figure}

\begin{figure}[htbp]
\includegraphics[scale=0.43, bb=-250 75 2000 650]{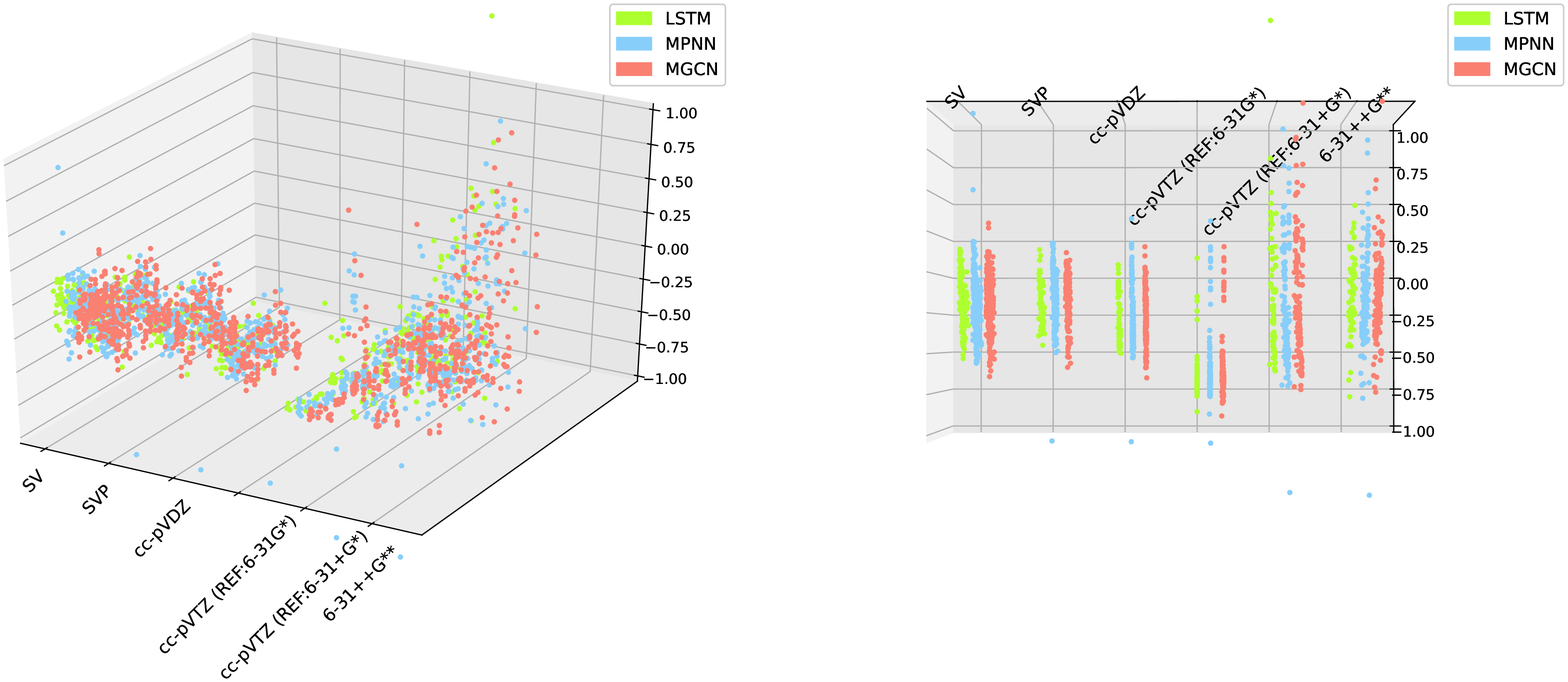}
\vspace{-1.25cm}
\caption{{Illustrations of the error scatters of predicted total CPU times using LSTM, MPNN, and MGCN models for the M06-2x cases in Fig.\ref{ref_notrained2}. Both the stereoscopic (left) the radial (right) distributions are shown.}} \label{ref_notrained2-ERR}
\end{figure}

{
For the case of either-trained predictions with the trained DFT functional and untrained basis set, the pre-trained models can be used to produce predictions with magnifications based on the fitted curves from various basis sets. This procedure was similar to that of untrained DFT functionals case, while it was a bit complex as the number of basis functions has a tight relationship with the actual computational times. 
To obtain the magnification between reference basis sets (REF-basis) and target basis sets (TAR-basis), a three-step procedure was used : 
1) choose the fitted curve of Eq.\ref{fbas} under the given DFT functional; 
2) obtain the $f^{bas}(x_{REF})$ and $f^{bas}(x_{TAR})$ based on the fitted curve of Eq.\ref{fbas};
3) calculate the magnification coefficient $c^{bas}$ based on Eq.\ref{fittedcurve}.
Once the $c^{bas}$ is obtained, the corrected timing predictions can be evaluated based on the predictions of reference basis sets.
Herein, the predicted total computational CPU times for 
several DFT functionals
together with basis sets of 6-31G, 6-31G*, 6-31+G*, 6-31++G**, SV\cite{schafer1992fully}, SVP\cite{schafer1992fully}, cc-pVDZ\cite{dunning1989gaussian}, and cc-pVTZ\cite{dunning1989gaussian} obtained using reference basis sets of 6-31G, 6-31G*, 6-31+G* were shown in Fig.\ref{ref_notrained2}. 
It can be seen that there were a lot of volatility of MREs of the original predictions that were based on the three reference basis sets (shown in the center of Fig.\ref{ref_notrained2}). 
The volatility in different reference models was caused by the enlarging magnification errors when target point and the reference point in fitted curve were too far from each other. These magnification errors can be largely reduced by introducing the similarity coefficients as we explained in the previous section. The similarity-corrected results were illustrated in the outside region of Fig.\ref{ref_notrained2} while the error scatters were shown in Fig.\ref{ref_notrained2-ERR} for the sampled M06-2x cases.
It can be seen that MREs can be largely reduced for all the four ML models. The graph-based MPNN and MGCN models behaved best, worse for the LSTM models, and worst for the RF models. 
One may notice that the relatively larger derivations for the cc-pVTZ case can be observed, it is because the 6-31G* basis set was used as the reference when applying the current similarity-corrected algorithm. The predicted timing results of cc-pVTZ basis set can be improved greatly when using 6-31+G* as the reference basis set. Additionally, one can also expect better predicted results if more basis sets can be used as the reference ones.
}

\begin{figure}[htbp]
\includegraphics[scale=0.75, bb=25 0 600 600]{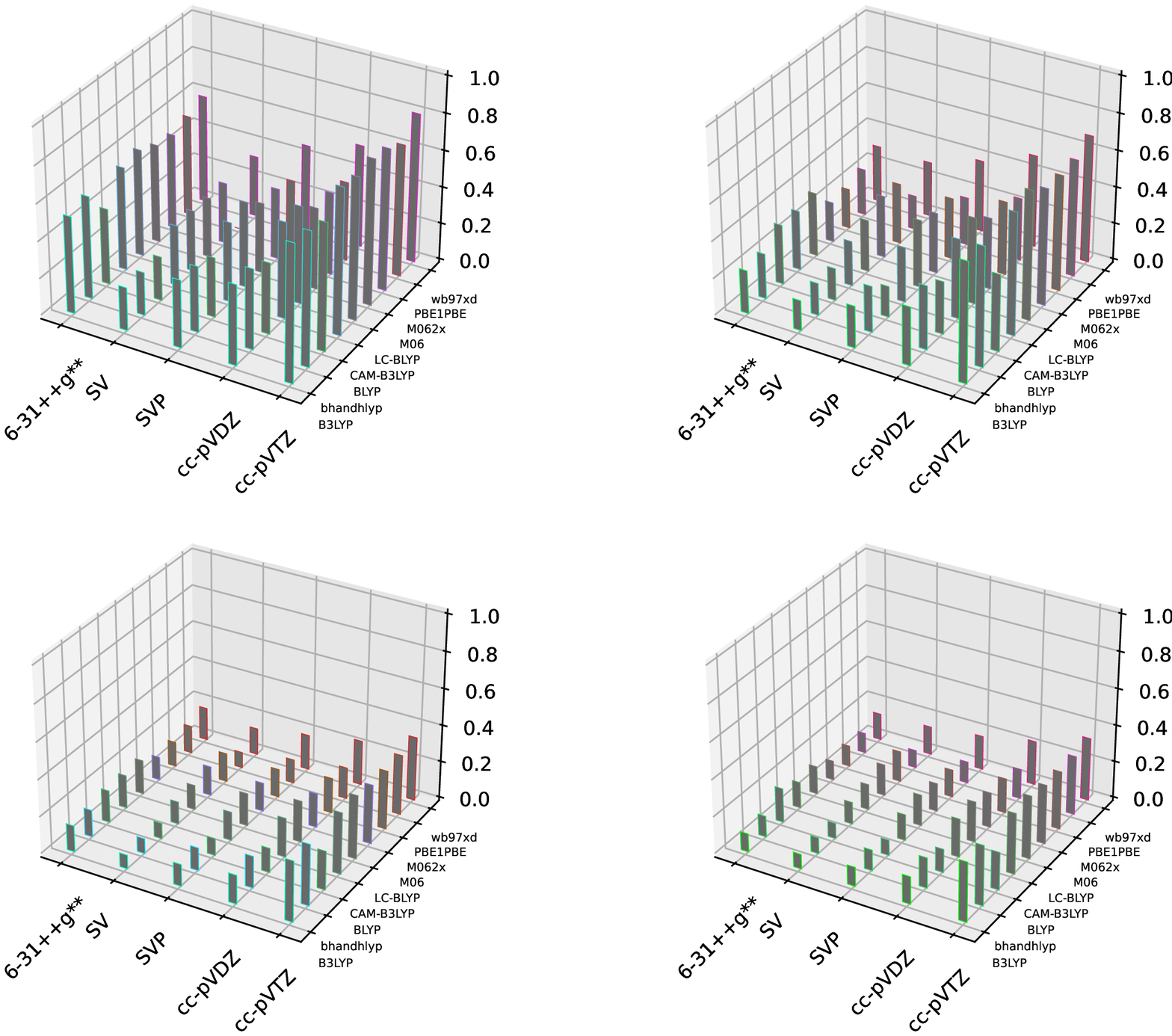}
\put(-420,470) {RF:}
\put(-150,470) {LSTM:}
\put(-420,250) {MPNN:}
\put(-150,250) {MGCN:}
\vspace{-3.0cm}
\caption{Illustrations of the MREs of predicted total CPU times with neither-trained models.}\label{ref_notrained3}
\end{figure}

\begin{figure}[htbp]
\includegraphics[scale=0.43, bb=-250 75 2000 650]{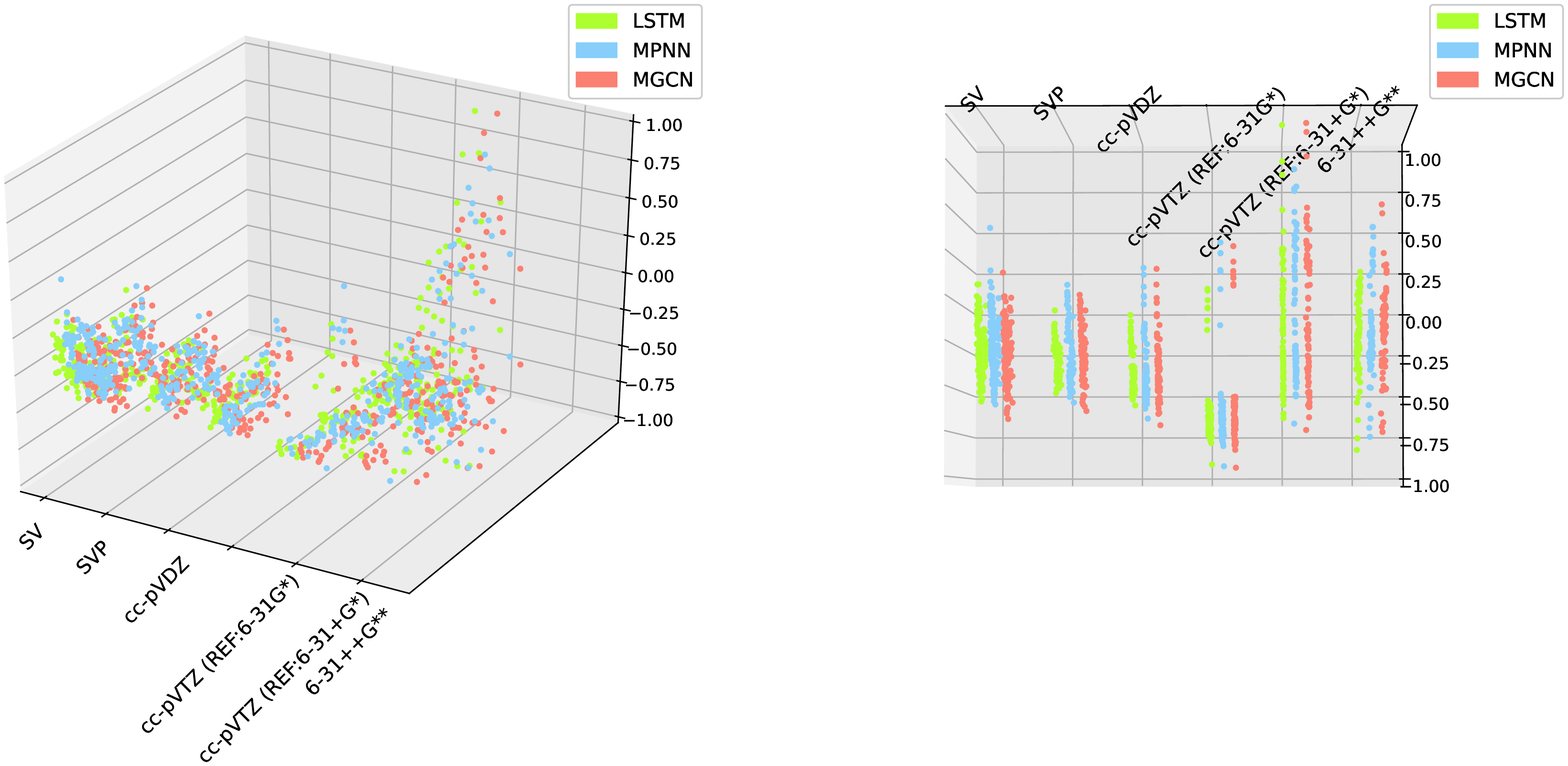}
\vspace{-2.5cm}
\caption{Illustrations of the error scatters of predicted total CPU times using LSTM, MPNN, and MGCN models (neither-trained) for the M06-2x cases in Figure.\ref{ref_notrained3}.} \label{ref_notrained3-ERR}
\end{figure}

{
For the case of neither-trained predictions with the untrained DFT functional and untrained basis set, the trained models can be used to produce predictions with magnifications both from the reference DFT functional and from the reference basis set under the chemical WMI anastz. 
Herein, the trained models of PBE/6-31G, PBE/6-31G*, and 6-31+G* combinations were used as the references, and total CPU times were predicted using the neither-trained models for several different combinations of DFT functionals and basis sets.   
The calculated MREs of predicted total CPU times were illustrated in Fig.\ref{ref_notrained3} and the error scatters were shown in Fig.\ref{ref_notrained3-ERR}. 
It can be noticed that the illustrated results were quite similar with those in Fig.\ref{ref_notrained2}. The graph-based MPNN and MGCN models still exhibited best, worse for the LSTM models, and worst behaviours for the RF models. Additionally, the reference basis set again played important role in the predicted results.
}

\section{Conclusion}

{A forecasting system for computational time of DFT/TDDFT calculations is presented in this work.
Four popular ML methods, including the RF, LSTM, MPNN, and MGCN models, were used to produce reliable predicted timing results for DFT/TDDFT calculations.
The structural similarity (in RF), recognition of chemical formula (in LSTM), and spatial structures (in MPNN and MGCN) were the ideas behind the choice of working ML models.
The cheminformatics with SMILES codes and graph-based spatial structures were employed for extracting structural information when training and testing the ML models. 
The use of cheminformatics also reduced the number of training suits significantly compared to that of the case when performing image recognition tasks.
Moreover, various combinations of DFT functional and basis set can be demonstrated by employing the proposed chemical MWI ansatz using either-trained or neither-trained models.}

{The four ML models can be used as the kernels for running the forecasting system. 
The overall performance followed the "LSTM $\rightarrow$ MGCN $\rightarrow$ MPNN $\rightarrow$ RF" order in the chemical space with the pre-trained ML models, and the typical MREs were 0.1 to 0.2 for the first two (LSTM/MGCN) models with +25\% to -25\% relative errors for most molecules. 
The tendency order in performance turned to "MGCN $\sim$ MPNN $\rightarrow$ LSTM $\rightarrow$  RF" in the chemical space without the pre-trained ML models and the typical MREs were still within the scope of 0.1 to 0.2 for the first two (MGCN/MPNN) models irrespective of whether either-trained or neither-trained cases was used. The distribution of relative errors was only slighted enlarged. The relatively small MREs and concentrated relative errors showed that the forecasting system can be used in the HPC's task assignment and load-balancing applications. 
We are currently working in this direction, particularly in coordination with fragmentation approaches in quantum chemistry, such as molecular fractionation with conjugate caps \cite{zhang2003molecular, he2006generalized, zhang2020gridmol2} and the renormalized exciton method\cite{al2005renormalized, ma2012new, ma2013calculating}.
}
 
{
At this stage, mainly the CPU time predictions on solving the Kohn-Sham equations as well as their TD expansions in single-point calculations are presented. This type of calculation is representative for the computational cost of a molecule, \cite{helgaker2014molecular} thus it is used as the first step in the design of a forecasting system. The routine calculations such as geometry-optimization, frequencies calculations are not yet included in these predictions, and are on the list of our future works. 
}
     
\section{Acknowledgement}

The work was supported by the National Key Research and Development Program of China (Grant No.2018YFB0203805), National Natural Science Foundation of China (Grant No.21703260), the Informationization Program of the Chinese Academy of Science (Grant No.XXH13506-403), and the Guangdong Provincial Key Laboratory of Biocomputing (Grant No.2016B030301007). We also thank PARATERA company for their cooperation.

\section{Statement}


Any data that support the findings of this paper is available from the corresponding author upon reasonable request.

\section{Appendix}

\subsection{Kohn-Sham density functional theory and its scaling}

In the Born-Oppenheimer approximation, a stationary electronic state can be described by a wave function $\Psi(\vec{r}_1,...,\vec{r}_N)$ satisfying the many-electron time-independent Schr\"{o}dinger equation,
\setcounter{equation}{0}
\renewcommand\theequation{A.\arabic{equation}}
\setcounter{figure}{0}
\renewcommand\thefigure{A.\arabic{figure}}    
\begin{equation}\label{eq_hami}
\hat{H} \Psi = \left[ \hat{T} + \hat{V} + \hat{U} \right] \Psi = \left[ \sum_{i=1}^{N} \left( -\frac{\hbar}{2m_i} \nabla_i^2 \right) +  \sum_{i=1}^{N} V(\vec{r}_i) + \sum_{i<j}^{N} U(\vec{r}_i,\vec{r}_j) \right] \Psi = E \Psi,
\end{equation}
\
where $\hat{H}$ denotes the electronic Hamiltonian, $E$ denotes the total energy, $\hat{T}$ denotes the kinetic energy, $\hat{V}$ denotes the potential energy from an external field due to positively charged nuclei, and $\hat{U}$ denotes the electron-electron interaction energy. 

In the KS-DFT hypothesis, particles can be treated as non-interacting fermions, so that there exists an orthogonal and normalized function set $\{\phi^{KS}_{i} |i=1,2,\cdots,N\}$ satisfying the condition:
\begin{equation}\label{eq_rho}
\rho (\vec{r}) = \rho_{s} (\vec{r}) = \sum_{i=1}^{N} {|\phi^{KS}_{i} (\vec{r})|}^2
\end{equation}
{Here,} $\rho (\vec{r})$ denotes the probability density of ground state electrons in a factual system and $\rho_{s} (\vec{r})$ denotes a fictitious system. 
Thus, the KS wave function is a single Slater determinant constructed from a set of function sets (i.e., orbitals) that are the lowest energy solutions to
\begin{equation}\label{eq_ks}
    [-\frac{\hbar}{2m_1}\nabla_1^2-\sum_{A}\frac{Z_A}{r_{1A}}+\int \frac{\rho(\vec{r_2})}{r_{12}}d\vec{r_2}+V_{XC}(1)]\phi_{i}^{KS}(1)=\varepsilon_{i}^{KS}\phi_{i}^{KS}(1),
\end{equation}
where the $V_{XC}(1)$ is referred to as exchange-correlation potential, 
and the "(1)" following each operator symbol simply indicates that the operator is 1-electron in nature.
This equation is very similar to the Fock equation,
\begin{equation}\label{eq_hf}
    [-\frac{\hbar}{2m_1}\nabla_1^2-\sum_{A}\frac{Z_A}{r_{1A}} - \frac{1}{2}\int \int \frac{\rho(\vec{r_1})\rho(\vec{r_2})}{r_{12}}d\vec{r_1}d\vec{r_2}]\phi_{i}^{HF}(1)=\varepsilon_{i}^{HF}\phi_{i}^{HF}(1),
\end{equation} 
{that is used in HF theory. 
Both Eq.\eqref{eq_ks} and Eq.\eqref{eq_hf} can be solved iteratively using the so-called self-consistent field (SCF) methods. 
During the SCF iteration, orbital $\phi_{i}$ is updated iteratively, and is used to calculate electron density , }
\begin{equation}\label{eq_rhovec}
    \rho(\vec{r})=\sum_{i=1}^{N}{|\phi_{i}(\vec{r})|}^2,
\end{equation}
which in turn, {determines} the {one-electron} matrix (e.g., Fock matrix in SCF iterations) to be diagonalized. 
After several iterations, both molecular orbital $\phi_i$ and its energy $\varepsilon_i$ can be obtained, {and the total electronic energy can then be calculated.}

Comparing Eq.\eqref{eq_ks} with Eq.\eqref{eq_hf}, one can clearly see that the major difference between them is in the {two-electron integrals component.} 
The origin of the $N^4$ scaling behavior is the calculation of four-center two-electron integrals, i.e.,
\begin{equation}\label{eq_2e}
    (\mu \nu | \lambda \sigma) = \int \int \phi_{\mu}(1) \phi_{\nu}(1) \frac{1}{r_{12}} \phi_{\lambda}(2) \phi_{\sigma}(2) d\tau_1 d\tau_2 , 
\end{equation}
where $\mu$, $\nu$, $\lambda$, and $\sigma$ denote indices of atomic orbitals. 
This scaling is also the upper boundary for the {HF} or hybrid DFT calculations.
However, many {two-electron} integrals are of negligible magnitude for large molecules, {and} some rigorous upper boundary conditions can be {applied} to the integrals. For instance, the Schwarz inequality \cite{steele2004cauchy}
\begin{equation}\label{eq_Schwarz}
    |(\mu \nu | \lambda \sigma)| \leqslant \sqrt{(\mu \nu | \mu \nu)(\lambda \sigma | \lambda \sigma)}
\end{equation}
allows the calculation of strict, mathematical upper bounds {of all two-electron} integrals to be computed in an $N^2$ $logN$ process. 
Apart from the calculation of the two-electron integrals, the diagonalization of the Fock or Fock-like matrix is expected to contribute significantly. 
{to the computational cost because the diagonalization step scales intrinsically as $N^3$; or even lower if the matrix (e.g., in large enough molecule) to be diagonalized is sufficiently sparse.}

Nevertheless, it can be noticed that for the hybrid DFT functionals, a hybrid exchange-correlation functional (i.e., the $V_{XC}(1)$ term in Eq.\eqref{eq_ks}) is usually constructed as a linear combination of the third terms (HF exact exchange functional) in Eq.\eqref{eq_hf}. Hence, hybrid DFT methods scale in a similar manner to the HF but {are} normally more expensive due to a larger proportionality term, involved while the pure DFT methods scale better than HF because there is no HF exchange.   

\subsection{ML models used in the forecasting system}

\subsubsection{RF model together with simple feed-forward neural networks}

\setcounter{equation}{0}
\renewcommand\theequation{B.\arabic{equation}}
\setcounter{figure}{0}
\renewcommand\thefigure{B.\arabic{figure}}

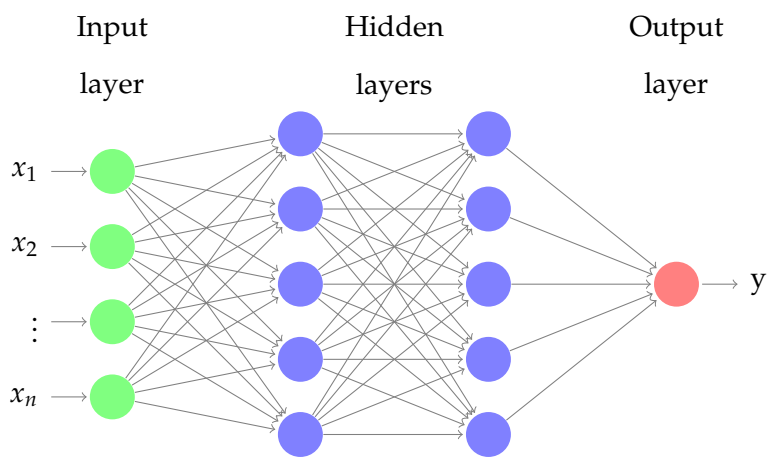
\begin{figure}[h]
\def\layersep{2.5cm}
\begin{tikzpicture}[shorten >=1pt,->,draw=black!50, node distance=\layersep]
    \tikzstyle{every pin edge}=[<-,shorten <=1pt]
    \tikzstyle{neuron}=[circle,fill=black!25,minimum size=17pt,inner sep=0pt]
    \tikzstyle{input neuron}=[neuron, fill=green!50];
    \tikzstyle{output neuron}=[neuron, fill=red!50];
    \tikzstyle{hidden neuron}=[neuron, fill=blue!50];
    \tikzstyle{annot} = [text width=4em, text centered]
    \node[input neuron,pin=left:{$x_1$}] (I-1) at (0, -1) {};
    \node[input neuron,pin=left:{$x_2$}] (I-2) at (0, -2) {};
    \node[input neuron,pin=left:{$\vdots$}] (I-3) at (0, -3) {};
    \node[input neuron,pin=left:{$x_n$}] (I-4) at (0, -4) {};
    \foreach \name / \y in {1,...,5}
        \path[yshift=0.5cm]
            node[hidden neuron] (H-\name) at (\layersep,-\y cm) {};    
    \foreach \name / \yy in {1,...,5}
        \path[yshift=0.5cm]
            node[hidden neuron] (HH-\name) at (2*\layersep,-\yy cm) {};
    \node[output neuron,pin={[pin edge={->}]right:y}] (O) at (3*\layersep,-2.5 cm) {};
    \foreach \source in {1,...,4}
        \foreach \dest in {1,...,5}
            \path (I-\source) edge (H-\dest);
    \foreach \source in {1,...,5}
        \foreach \dest in {1,...,5}
            \path (H-\source) edge (HH-\dest);        
    \foreach \source in {1,...,5}
        \path (HH-\source) edge (O);
    \node[annot] at (0,0.5 cm) {Input layer};
    \node[annot] (hl) at (\layersep+1.25 cm,0.5 cm) {Hidden layers};
    \node[annot] at (\layersep+5 cm,0.5 cm) {Output layer};
\end{tikzpicture}
\caption{The simple FNN model.} \label{neuro}
\end{figure}

We used the feed-forward neural network (FNN) as the skeletal frame to obtain the model between basis number and the computational time. FIG.\ref{neuro} shows an illustration of the FNN model. 
Four layers are used in our model, which are input layer, two hidden layers, and the output layer. 
The "input layer" is constructed using the system magnitude features (e.g., number of basis sets). 
These vectors are normalized and then fed to the hidden layers. 
Each "hidden layer" contains several neurons, and the {\sc tanh} function is used as the activation function. 
The data passed from the hidden layer will be directly used in the linear combination of the output results.

\begin{figure}
	\centering
	\includegraphics[scale=0.65,bb=500 0 200 300]{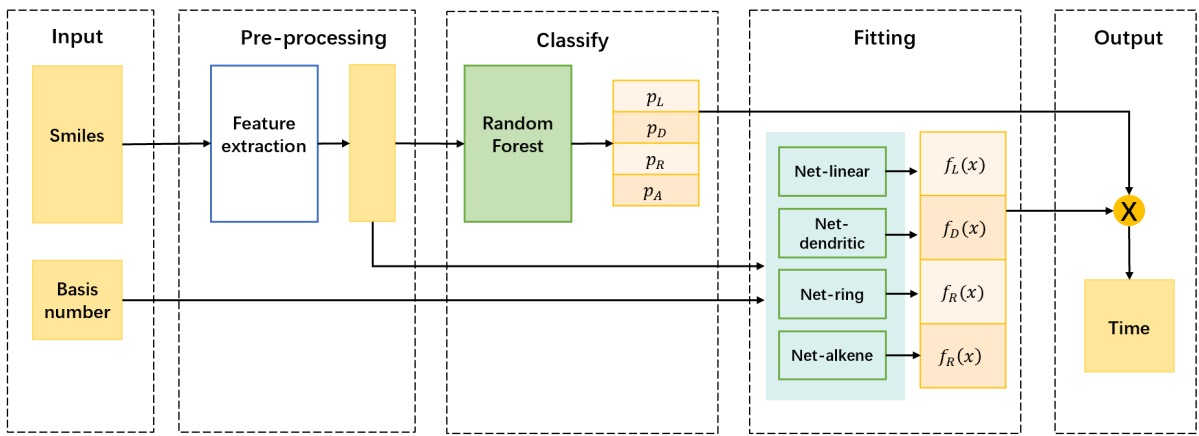}
	\caption{Architecture of RF with FNNs.}
	\label{rfarc}
\end{figure}

The predicted result for a molecule that is far from the training set and will still be very poor, if we only use the FNN model of FIG.\ref{neuro},  
To overcome the dependency of training sets, we introduced the idea of "feature training". 
"Feature training" means that few molecules suites with specific features (e.g. linear(L), dendritic(D), ring(R), etc.) will be trained as the "cost functions", 
upon which the computational cost ($y$) for any molecule can be calculated by a linear combination of these "cost basis", e.g., 
\begin{equation}\label{eq_cost}
    y = p_L \cdot f_L(x) + p_D \cdot f_D(x) + p_R \cdot f_R(x) + ... ,
\end{equation}
where $p_L$, $p_D$ {and} $p_R$ denote the possibilities for each "cost basis" $f_{feature}(x)$, the $f_{feature}$ denotes the "cost functions", and the $f_{feature}(x)$ denotes the expected computational cost for the "feature" model with magnitude parameter $x$. Herein, one can notice that this "feature training" ansatz matches well with the RF model used in ML.
Under this ansatz, a specific model (i.e. cost function) is trained and saved for each structure of the molecular suit. 
Afterwards, an RF classifier is used to place the molecules into given categories, such as linear, dendritic or ring molecules, etc., with possibilities ($p$). 
The classifier accepts the {\sc SMILES} codes of molecules as its input. 
The number of every molecule's atoms (with hydrogen atoms excluded), branches, atoms on branches and cycles is calculated and combined as an input vector according to its {\sc SMILES} code.
Fig.\ref{rfarc} illustrates entire process. 

\textbf{In training :}

The coefficients for each "cost basis" were obtained via the RF classifier with the {\sc scikit-learn(sklearn)}\cite{scikit-learn} package, with all the parameters set to default values. The classifier provides a molecule's probability of falling into each category as the output, which can be used in Eq.(\ref{eq_cost}) for the predictions.

\subsubsection{Bi-LSTM with attention}

The kinds of features extracted by the RF classifier are designed artificially with subjective preference,
so that it may be not enough for aggregating molecular structural information. 
Considering that we used textual data (i.e., SMILES code) as the representation of molecular structure, methods for natural language processing (NLP) {are thus, suitable for feature extraction as a result of this issue.} 

Here, we use the bidirectional LSTM(Bi-LSTM) with attention model that was proposed by Peng and {his} coworkers.\cite{Peng2016Attention} It is the state-of-the-art model for relation classification tasks; thus, we used this model to extract structural features from SMILES.
Fig.\ref{lstm} illustrates the architecture of the model. The input features include the SMILES code (in the form of one-hot) and the number of basis functions.  
Suppose every character in a T-length SIMLES sequence is denoted with a one-hot vector $x_i$, $x_i$ will be converted to a real number vector $e_i$ ($e_i=Wx_i$, {where $W$ denotes a parameter matrix automatically learned by the model during training}). Then $E=\{e_1,e_2,\cdots ,e_t\}$ is sent to the Bi-LSTM layer. The Bi-LSTM layer consists of a forward  layer and a backward layer so that the model can learn from forward and backward sequences {as} the past and future semantic information in a sentence are equivalently significant. The LSTM layers at two directions generate two outputs, $H_f$ and $H_b$:
\begin{equation}
H_f=L(E),
\quad H_b=L(E),
\end{equation}
where $L$ denotes the operations performed by a LSTM layer.
Then an attention layer accepts the sum of the outputs from Bi-LSTM layers:
\begin{equation}
H=H_f+H_b.
\end{equation}

{The attention mechanism allows different context vectors to be generated from the Bi-LSTM layer's output at every time step by assigning different "attention weight" to the outputs. 
Without the attention mechanism, the feature extraction operation on the output at every time step would have the limitation of depending on one same context vector with fixed length invariant to time steps.   
The attention layer outputs the final representation $c$ of a SMILES as} 
\begin{equation}
\alpha =softmax(w^T tanh(H)),
\end{equation}
\begin{equation}\label{eq_c}
c=H {\alpha}^T,
\end{equation}
where $\alpha$ denotes the attention weight vector, and $w$ denotes a trained parameter vector. The high-level features of molecular structures are produced after the attention layer. We combine the structural features and the number of basis functions and feed them into fully-connected layers to get the predicted result.

\begin{figure}
\centering
\label{lstm}
\includegraphics[scale=0.14, bb=00 0 4000 1000]{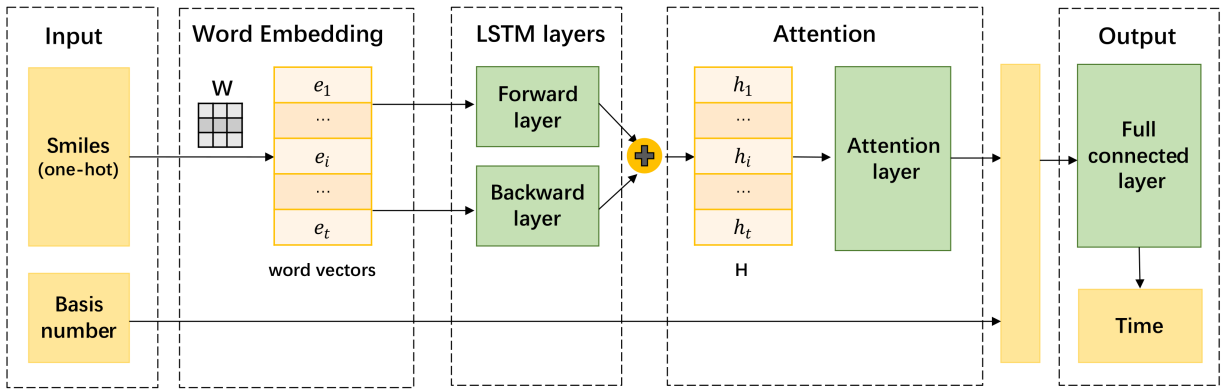}
\caption{Architecture of Bi-LSTM with attention model.}
\label{lstm}
\end{figure}

%

\subsubsection{MPNN model}     

\begin{figure}
	\centering
	\includegraphics[scale=0.14,bb=0 0 4000 1000]{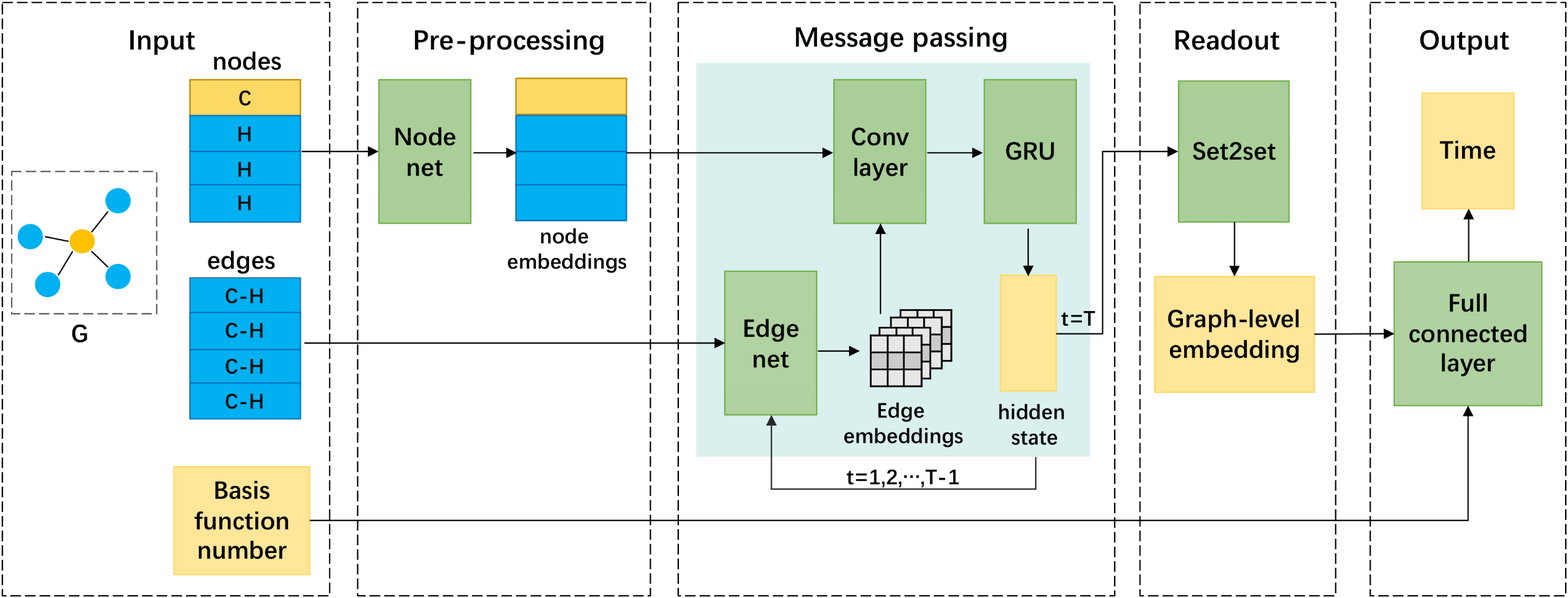}
	\caption{Architecture of the MPNN model.} \label{mpnn}
\end{figure}

As a representation of molecular structure, SMILES is quite {inexact} due to absence of spatial information. Fig.\ref{mpnn} shows that for a more accurate representation, it is rational to model a molecule using an undirected graph \textit{G}. We use the MPNN model, which is recognized as a kind of graph neural network (GNN) proposed by Gilmer as a solution for graph-based learning.\cite{Gilmer2017Neural}  

The initial inputs of the model include a feature vector collection for nodes of the graph, denoted with $x_v$, containing features of atom types, aromaticity and hybridization types, and a feature vector collection for edges, denoted with $e_{vw}$, containing features of bond types. {Further, the model has} two phases: a message passing phase and a readout phase. The message passing phase runs \textit{T}-step graph convolutions and at each step \textit{t}, it is defined in terms of a message function $M_t$ and a vertex update function $U_t$. Before the message passing, the node vectors are mapped to a $n \times d$ matrix called "node embedding" by a network (called "node net"), with $n$ {denoting the number of nodes; and $d$ representing the dimension of hidden state of each node. During the message passing phase, hidden states $h^t_v$ of each node are updated according to messages $m^{t+1}_v$. The message passing phase can be summarized as}
\begin{equation}\label{eq_c}
 m^{t+1}_v=\sum_{w \in N(v)} M_t(h_v^t,h^t_w,e_{vw}), 
\end{equation}
\begin{equation}\label{eq_c}
 h_v^{t+1}=U_t(h_v^t,m_v^{t+1}), 
\end{equation}
where \textit{N(v)} denotes the neighbours of \textit{v} in \textit{G}. $M_t$ is defined as $M(h_v,h_w,e_{ew})=A(e_{vw})h_w$ specifically, where $A(e_{vw})$ denotes a network (edge net) mapping each edge vector $e_{vw}$ to a $d \times d$ matrix (edge embedding). The vertex update function is GRU, short for gated recurrent unit.\cite{cho2014On} At the readout phase, a feature vector can be obtained as a summary of the whole graph with a readout function \textit{R}
\begin{equation}\label{eq_c}
 \hat{y}=R(\{h_v^T |v \in G \}),   
\end{equation}
where \textit{R} denotes the \textit{set2set} \cite{Vinyals2016Order} model. The \textit{set2set} model produces a graph-level embedding that is invariant to the order of nodes. Finally, we combine the graph-level embedding and the basis function number and feed them to fully-connected networks to get the prediction results. Fig.\ref{mpnn} illustrates the architecture of the MPNN model.

%
 
\subsubsection{MGCN model}

\begin{figure}
	\centering
	\includegraphics[scale=0.14,bb=00 0 4000 1000]{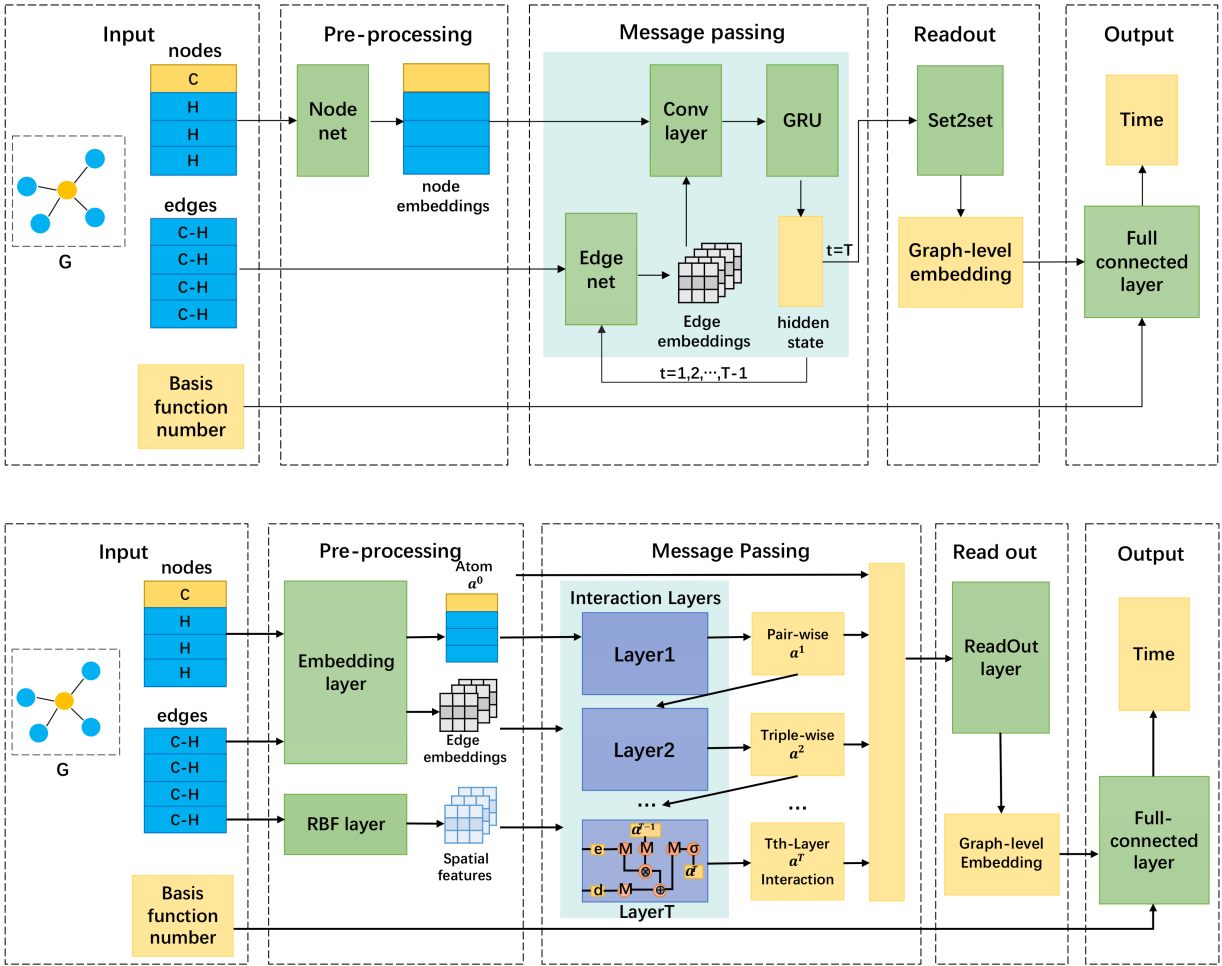}
	\caption{Architecture of the MGCN model.} \label{mgcn}
\end{figure}

Apart from MPNN, we also introduced another GNN model, MGCN\cite{Lu2019Molecular}, which is reported to have the advantages of generalizability and transferability.
As shown in Fig.\ref{mgcn}, the architecture of MGCN includes five phases. 
First, the initial inputs of the model include a feature vector collection for nodes of the graph, denoted by $a^0\in \mathbb{R}^{D}$, containing features of atom types, aromaticity and hybridization types, a feature vector collection for edges, denoted by $e\in \mathbb{R}$, containing features of bond types and bond lengths, and the number of the basis functions. At the pre-processing phase, the embedding layer generates the node atom embeddings ($A^0\in \mathbb{R^{N \times D}}$) and edge embeddings ($E \in \mathbb{R^{N \times N \times D}}$). The radial basis function (RBF)\cite{broomhead1988radial,schwenker2001three} layer converts the bond lengths to a distance tensor $D \in \mathbb{R}^{N \times N \times K}$, with $d_{ij}$ representing the distance between atom \textit{i} and atom \textit{j}. The RBF layer's function form {can be expressed} as
\begin{equation}\label{eq_rbf}
RBF(x)=\mathop{\frown}\limits_{i=1}^Kh(||x-\mu_i||),
\end{equation}
where $\frown$ denotes concatenation and $\mu_i$ is from a set of K central points $\{\mu_1,\dots,\mu_K\}$. 
At the message passing phase, the interaction layers are constructed in the form of hierarchical architecture to simulate the many-body interactions which are transformed at different levels (atom-wise, atom-pair, atom-triple, etc.). The $l$-th layer generates an edge presentation $e^{l+1}_{ij}$ and an atom representation $a^{l+1}_i$:
\begin{equation}\label{eq_eij}
e_{ij}^{l+1}=h_e(a_i^l,a_j^l,e_{ij}^l),
\end{equation}
\begin{equation}\label{eq_ai}
a_i^{l+1}=\sum\limits_{j=1,j\ne i}^N=h_v(a_j^l,e_{ij}^l,d_{ij}),
\end{equation}
where $h_e$ denotes the edge update function and $h_v$ denotes the message passing function, respectively. The form of $h_e$ is {as follows}:
\begin{equation}\label{eq_he}
h_e=\eta e_{ij}^l \oplus (1-\eta)W^{ue}a_i^l \odot a_j^l.
\end{equation}
Here $\eta$ denotes a constant set to 0.8, $W^{ue}$ denotes a weight matrix, $\oplus$ denotes {elementwise addition and $\odot$ is the elementwise dot product. }
The form of $h_v$ is
\begin{equation}\label{eq_hv}
h_v=tanh (W^{uv}(M^{fa}(a_j^l)\odot M^{fd}(d_{ij}) \oplus M^{fe}(e_{ij}) )),
\end{equation}
where  \textit{M(x)} denotes a linear layer which is in the form $M(x)=Wx+b$. The outputs of T interaction layers along with $a_i^0$ are concatenated together as 
\begin{equation}\label{eq_ai2}
a_i=\mathop{\frown}\limits_{k=0}^Ta_i^k. 
\end{equation}
Afterwards, The Readout layer generates a graph-level embedding G as
\begin{equation}\label{eq_G}
G=\sum \limits_{i=1}^{N}W^{r_2^a}\sigma (M^{r_1^a}(a_i))+\sum\limits_{i=1}^N\sum \limits_{j=1,j\neq i}^NW^{r_2^e}\sigma(M^{r_1^e}(e_{ij})).
\end{equation}
Here $\sigma$ denotes the softplus function. Finally, G and the basis function number are concatenated and sent to a fully-connected layer to get the prediction time.

\subsection{Parameters when training the models}

The number of samples in training  and testing sets, as well as some of important parameters in the training process were listed in Tab.\ref{para}. 

\setcounter{table}{0}
\renewcommand{\thetable}{C\arabic{table}} 
\begin{table}[h]	
	\caption{The number of samples in training ($N^{Train}_{Samples}$) and testing ($N^{Test}_{Samples}$) sets, and some of important parameters ($BatchSize$, $StepSize$) in training models. } 
	\label{para}	
	
	\begin{tabular}{cccccccccccccccccccccccc}
		\hline
		\hline
	Data        & & $Model$ & $N^{Train}_{Samples}$ &  $N^{Test}_{Samples}$  &  $BatchSize$ &  $StepSize$   \\ 
		\hline		
\multirow{4}*{Tab.1 and Fig.7}	& &	   RF   &   108  &  25  &    40  &  0.005   \\
                                & &	  LSTM  &   500  &  25  &    100  &  0.005   \\
                                & &	  MPNN  &  1500  &  25  &    50  &  0.005   \\
                                & &	  MGCN  &   500  &  25  &    25  &  0.001   \\
		\hline
\multirow{3}*{Tab.II}           & &   LSTM  &  100-1500  &  25  &    100  &  0.001-0.005   \\
	                            & &   MPNN  &  100-1500  &  25  &    100  &  0.001-0.005   \\
	        	                & &   MGCN  &  100-1500  &  25  &    100  &  0.001-0.005   \\
		\hline                  
                                & &   RF    & 1478  &  238  &    40  &  0.005  \\
    {Fig.9-11}                  & &   LSTM  & 1478  &  238  &   100  &  0.005   \\  
   {(Fig.14-19)$^a$}            & &   MPNN  & 1478  &  238  &   100  &  0.001  \\
                                & &   MGCN  & 1478  &  238  &   100  &  0.005   \\
 		\hline                                  
	\end{tabular}

\footnotesize{$^a$ The trained models were re-used for predicting the results with $N^{Test}_{Samples}$ is 49 in Fig.14-19.}\\
\end{table}

%

\renewcommand\refname{References} 
\bibliography{era_prediction}
\end{document}